\documentclass[
  aps,
  prl,
  twoside,
  twocolumn,
  showpacs,
  floatfix
]{revtex4}

\usepackage{amsmath}
\usepackage{amssymb}
\usepackage{graphicx}
\usepackage{tabularx}
\usepackage{dcolumn}
\usepackage{longtable}
\usepackage{footnote}

\begin{document}

\newcommand{\bec}{\begin{center}}
\newcommand{\ec}{\end{center}}
\newcommand{\be}{\begin{equation}}
\newcommand{\ee}{\end{equation}}
\newcommand{\beqn}{\begin{eqnarray}}
\newcommand{\eeqn}{\end{eqnarray}}
\newcommand{\bet}{\begin{table}}
\newcommand{\ent}{\end{table}}
\newcommand{\bib}{\bibitem}

\newcommand{\sect}[1]{Sect.~\ref{#1}}
\newcommand{\fig}[1]{Fig.~\ref{#1}}
\newcommand{\Eq}[1]{Eq.~(\ref{#1})}
\newcommand{\eq}[1]{(\ref{#1})}
\newcommand{\tab}[1]{Table~\ref{#1}}

\renewcommand{\vec}[1]{\ensuremath\boldsymbol{#1}}
\renewcommand{\epsilon}[0]{\varepsilon}

\newcommand{\cmt}[1]{\emph{\color{red}#1}%
  \marginpar{{\color{red}\bfseries $!!$}}}



\title{
The DFT and molecular dynamics multiscale study of the corrugation of graphene on Ru(0001):
the unexpected stability of the moire-buckled structure
}


\author{P. S\"ule} \email{sule@mfa.kfki.hu} 

\affiliation
{Research Centre for Natural Sciences,
Institute for Technical Physics and Materials Science
\\
Konkoly Thege u. 29-33, Budapest, Hungary,sule@mfa.kfki.hu\\
}

\date{\today}


\begin{abstract}
Results from first principles density functional theory (DFT)
calculations and classical molecular dynamics (CMD) simulations
are presented on moire-corrugation of graphene (gr).
 We find that the moire-corrugated graphene could be surprisingly stable against the perfectly flat gr-sheet
as pointed out by CMD simulations and DFT calculations.
We also show that using the cost-effective
CMD approach one can simulate graphene
on e.g. Ru(0001) with a correct binding registry and reasonable corrugation
and adhesion energy.
A new force field has been parameterized for the interface using an
angular-dependent
Abell-Tersoff potential.
The newly parameterized Abell-Tersoff interface potential provides correct moir$\acute{e}$ superstructures
in accordance with scanning tunnelling microscopy images and with
DFT results.
Based on {\em ab initio} DFT calculations,
we also find that the CMD moir$\acute{e}$ superstructure
can be used as a preoptimized structure for DFT calculations
and for further geometry optimization.
The nearly flat gr (the corrugation $\xi \approx 0.2$ $\hbox{\AA}$) on Ru(0001) is slightly energetically unfavorable vs. the moire-corrugated
gr-system ($\xi \approx 2.0$ $\hbox{\AA}$) as revealed by van der Waals DFT structural relaxation.

\end{abstract}

\maketitle

\section{Introduction}

 The surface topography of the graphene (gr) film
induced by the interfacial interaction with the substrate
could be a key factor in the change of the fine properties of the
supported gr-film \cite{Batzill,rev,gr}.
 Recent studies reveal that, the extrinsic morphology of supported graphene 
is controlled (periodically repeated moir$\acute{e}$ superstructures), distinct from the random intrinsic (rippled) morphology of freestanding
graphene \cite{rippledgr}.
Monolayer graphene on various substrates has been shown to alter also from the ideally 
flat layer form \cite{Batzill,rev} and nearly flat (ultraflat and ripple-free) graphene
has been found on few substrates only \cite{ultraflatgr_sio2,ultraflatgr_bn}.
Graphene promises a broad range of applications in its nearly flat
form, however, unexpected distorsions in the sheet
could break down its remarkable  properties \cite{gr,Tapaszto}.
Therefore, it is important to understand how to controll
the morphology of gr layers either transferred from other supports
or grown on various substrates.
Recent experimental and theoretical studies have shown
that gr could be seriously buckled when placed on rigid substrates
depending on the strength of gr-substrate interaction \cite{Batzill,rev}.


 Although first principles calculations (such as density functional theory, DFT) have widely been
used in the last few years to understand corrugation of nanoscale gr sheets
on various substrates \cite{DFT:Ru-Hutter,DFT:Ru-Stradi,Peng,DFT:Ru_Wang,DFT:Ru_corrug}
the modelling of larger systems above $1000$ Carbon atoms
remains, however challenging.
The simulation of large scale systems can in principle be done by classical
molecular dynamics simulations.
However, until now, the results of simulations of supported gr layer
have been reported in few cases only for SiC support \cite{gr-md}.
Results are also published very recently for gr/Cu(111) moire superstructures
using CMD simulations \cite{gr-md-cu}.
In another study the binding of gr to Cu(111) has been studied
by a simple pair potential for the interface \cite{gr-md-cu2}.
The reason of the very few reports could be the difficulty of the adequate description
of the weakly bounded gr/substrate interface by the
available simple pairwise interfacial force fields.
The inclusion of proper bond orientations at the interface via
angular-dependent gr-substrate potentials is still lacking.
We will show that
simple pairwise potentials tend to flatten gr and 
favor improper binding sites because of the exclusion of
adequate angular orientations at the interface.
New parameter sets have been developed for the angular-dependent
Abell-Tersoff potential \cite{Abell,Tersoff} at the interface which describe
gr-Ru(0001) bonding adequately.


 We would like to show that using a newly
parameterized interfacial Abell-Tersoff (AT) potential \cite{Abell,Tersoff} 
one can account for the observed surface reconstructions of gr.
 The experimentally seen
superstructures and corrugation for gr/Ru(0001) are reproduced for the first time then by classical
molecular dynamics simulations using the new parameterized AT potential
at the C-Ru interface.
The new parameter set could open new way to large scale simulations
on various properties of gr/Ru(0001) system.
The CMD approach could provide corrugated starting geometry for
expensive DFT calculations.
To demonstrate this,
van der Waals-type {\em ab initio} DFT calculations together with
accurate geometry optimization have also been carried out for the
CMD relaxed moire-corrugated structures which reveal some
unexpected features.

\section{Methodology}


 Classical molecular dynamics has been used as implemented
in the LAMMPS code (Large-scale Atomic/Molecular Massively Parallel Simulator) \cite{lammps}.
The graphene  layer has been placed incommensurately on the substrates.
The commensurate displacement is energetically
inaccessible (unfavorable), since on-top positions are available
only for a certain part of Carbon atoms when lattice
mismatch exceeds a critical value ($\sim 2-3$ $\%$) \cite{DFT:Ru-Hutter}.
The proper lateral adjustment of C-C bonds would require too much
strain which is clearly unfavorable for most of the
gr-support systems \cite{DFT:Ru-Hutter}. 
The graphene atoms form then an incommensurate overlayer by occupying
partly registered positions (hexagonal hollow and top sites) which
leads to a moir$\acute{e}$ superstructure
(long-wave coincidence structures) \cite{rev}.

 The molecular dynamics simulations allow the optimal lateral
positioning of the gr layer in order to reach nearly epitaxial displacement
and the minimization of lattice misfit.
The relaxation of the systems have been reached in two steps:
first geometry optimization has been carried out and then
CMD simulations have been utilized in
few tens of a thousand simulation steps.

{\em Empirical potentials}:
 For the graphene layer 
the Adaptive Intermolecular Reactive Empirical Bond Order (AIREBO) potential 
\cite{stuart} and the Albe-Erhart \cite{SiC} bond-order potential have been used.
For the Ru substrate, a recent EAM \cite{EAM:Ru} potential is utilized.

 The interaction between the graphene layer and the substrate
has been modelled first by a simple Morse pairwise potential.
\be
E_{Morse}=D_0 \biggm[ e^{-2 \alpha (r-r_0)}-2 e^{-\alpha (r-r_0)} \biggm],
\ee
where $D_0$, $\alpha$ and $r_0$ are the parameters to be fitted.
For most of the systems we find that $\alpha=5.5$ $\hbox{\AA}^{-1}$ is an appropriate choice.
The rest of the parameters has been varied in order to
reproduce the DFT results available in the literature \cite{Batzill,rev}.
For instance, for the gr-Ru(0001) system $D_0=0.05$ and $r_0=2.1$.
Using these values the interaction energy  is around $\Delta E \approx -0.2$
eV/C atom which is typically in the range of weak chemical or strong physical adsorption. 
  Lennard-Jones (LJ) potential (6-12) has also been used for the interface
as implemented in LAMMPS \cite{lammps}.

\subsection{Angular dependence and topography}

 The simple Morse or Lennard-Jones pairwise potentials although provide a relatively adequate 
representation of moir$\acute{e}$ superstructures in many respect, however,
one important aspect of the topography is wrong.
As we realized,
Morse and other simple pair potentials overbind hollow sites and
valleys (hollow-bumps) occur instead of hollow-humps as it is shown
in Figs 1(a) and 1(c).
DFT studies favor to bind the hump-like protrusions to hollow sites \cite{DFT:Ru-Hutter}.
To overcome this problem we introduce angular dependency in the pair potential
using Abell-Tersoff potential \cite{Abell,Tersoff} for the gr-substrate interaction.
Proper bond-angle orientations at the interface could help to
stabilize hollow-bumps vs. hollow-humps.

  The Albe-Erhart (AE) parameterization of C-C
were used for the gr-sheet \cite{SiC} when combined
with the AT-interface.
The AE potential can easily be transformed into AT potential \cite{pontifix,SiC}
since esentially the angular part is the same only the radial part 
is different, although they are fully equivalent \cite{pontifix,SiC}.

 Originally, the Abell-Tersoff \cite{Abell,Tersoff} potential is given in the following form:
\be
V_{Tersoff}=\sum_{ij,i>j} f_{ij}(r_{ij})[V_{ij}^R-b_{ij}(\Theta) V^A_{ij}(r_{ij})].
\ee
The radial part of the AT potential is composed of the
following repulsive and attractive functions,
\be
V_{ij}^R=A exp(-\lambda_1 r),
\ee
\be
V_{ij}^A=B exp(-\lambda_2 r).
\ee
The angular dependence is introduced via the attractive part $V_{ij}^A$ term by
the $b_{ij}(\Theta)$.
\be
b_{ij}(\Theta)=(1+\beta^n \chi_{ij}^n(\Theta))^{-\frac{1}{2n}}
\ee

\be
\chi(\Theta)=\sum_{k(\neq i,j)} f_{ik}^c(r_{ik}) g_{ik}(\Theta_{ijk})  exp[2\mu_{ik}(r_{ij}-r_{ik})]
\ee
where the angular term $g(\Theta)$,
\be
g(\Theta)=\gamma \biggm(1+\frac{c^2}{d^2}-\frac{c^2}{d^2+(cos\Theta-h)^2}\biggm),
\ee
where $h=-cos(\Theta_0)$. We find that $\Theta_0 \approx 80^{\circ}$ is the
most favorable C-C...Ru bond angle for graphene.
The cutoff function is
\[
f_{ij}(r_{ij})= \left\{ \begin{array}{cc}
~~~~~~~~~ 1 & r \le R_c-D_c  \\ \frac{1}{2}-\frac{1}{2}sin[\frac{\pi}{2}(r-R_c)/D_c] & |r-R_c| \le D_c
 \\
0 & r \ge R_c+D_c
\end{array} \right. \]
where $R_c \le 3.0$ $\hbox{\AA}$ is the short range cutoff distance
and $D_c=0.2$ $\hbox{\AA}$.

 Bond angle analysis of the relaxed interface provides
C-C-Ru bond angles in the range of $\Theta_0=80^{\circ} \pm 5^{\circ}$.
We find that within this range of $\Theta$
the obtained pattern is the expected one with periodic humps-and-bumps.
Outside this regime the surface is decorated with
other unexpected features (more complicated pattern in the inter
bump region, or distorted bumps).
 Introducing this term one can influence the energetic stability of various
conformations as well as the the most favorable surface topography.
Neglecting e.g. this term, or setting in $\Theta_0 > 90^{\circ}$
hollow sites become the most favorable binding sites which is
not confirmed by DFT calculations \cite{DFT:Ru-Hutter,DFT:Ru-Stradi,DFT:Ru_Wang,DFT:Ru_corrug}.

\subsection{The fitting procedure}

 The fitting of the angular and radial parts of the interface potential were carried out on a limited data base
which includes first neighbor distances and adhesion energy of
fictitious weakly bound dimer, B1 (NaCl), B2 (CsCl) and B3 (ZnS) alloys of C and Ru.
The traditional way of fitting procedure (see e.g. refs. \cite{SiC,GaN})
does not work in this special case when an interface potential is to be
parameterized.
In a standard situation one should fit the AT function to the
experimental lattice constants, cohesive energies and bulk moduli of various polymorphs
of RuC \cite{RuC}.
However, in this case the interface potential would bind graphene
too strongly to Ru(0001) (chemical adhesion).
Photoelectron
spectroscopy shows that the layer bonding is not carbidic \cite{Parga}.
The bonding situation and the chemical environment is completely
different in gr/Ru(0001) and in RuC.
Even if a weak chemical bonding takes place in gr/Ru(0001), it is far much weaker
than in RuC \cite{DFT:Ru-Hutter,DFT:Ru-Stradi,DFT:Ru_Wang,DFT:Ru_corrug}.
Using a RuC based fitted potential
the adhesion energy of gr/Ru(0001) would be $E_{adh} \gg 1$ eV/C, which is far higher than
the measured and the DFT calculated $E_{adh} \approx 0.2$ eV/C \cite{DFT:Ru-Hutter,DFT:Ru-Stradi,DFT:Ru_Wang,DFT:Ru_corrug}.
Moreover, in the RuC dimer molecule, e.g. the dissociation energy is in the range of
6.6 eV \cite{RuCdim} which is far above the adhesion energy of
gr/Ru(0001) of 0.2 eV/C \cite{DFT:Ru-Hutter}, hence the
dimer properties neither can be used for fitting.
Moreover, the equilibrium distance of the RuC dimer is also too short ($1.66$ $\hbox{\AA}$
when compared with the expected $\sim 2.1$ $\hbox{\AA}$ distance of the
gr/Ru(0001) complex.
Hence one can not use the available experimental data set of RuC for parameterization.

 Therefore one has to choose a different strategy for fitting the Tersoff
parameters.
We find that simple (fictitious) alloys with proper bond distances and cohesive energies  are sufficient for
efficient parameterization.
The training set is necessary only to parameterize
the Tersoff function for reproducing
the weak adhesion energy and bonding distance of C-Ru complexes
at a proper C-C-Ru bonding angle.
In particular, in the fitting data base (training set)
we use the bond distance $r_0=2.1$ $\hbox{\AA}$
and for the adhesion energy $0.2$ eV/C which has been found by
DFT for gr/Ru system \cite{DFT:Ru-Hutter,DFT:Ru-Stradi,DFT:Ru_Wang,DFT:Ru_corrug}.
The obtained parameter set is to be refined on a trial-and-error basis
on small or medium sized gr/Ru(0001) systems (5000-10000 C atoms on
10 layers Ru(0001)).
Using few steps (in each step with new parameterization procedure
for the angular part and for the other free parameters in the
radial part)  
one can adjust in principle the interfacial Tersoff function
in order to reproduce the most important properties of gr/Ru(0001)
(corrugation, C-Ru bonding energy, adhesion energy, supercell,
proper binding registry).

 Parameters in the angular term ($\gamma$, c,d) are to be determined then by
least square fitting procedures such as installed in the code PONTIFIX \cite{pontifix,GaN,SiC}. 
Parameter $h$ ($-cos(\Theta)$) has been kept fixed in order to
force the C-C...Ru bond angle into the preferred $\sim 80^{\circ}$ value.
First, we fit the radial part of the Tersoff function (which is a fully equivalent
Morse-like potential) to the Morse function used for previous simulations.
We use this set of radial parameters as an initial guess for
the parameterization of the full angular-potential.
In particular, we utilize the also fully equivalent 
Morse-like functions used in the Brenner bond-order potential and which
has been implemented in the PONTIFIX code \cite{Tersoff-Brenner,GaN,SiC}.
\be
V_{ij}^R=\frac{D_0}{S-1} exp(-\beta_{AE} \sqrt{2S}(r-r_0))
\ee
\be
V_{ij}^A=\frac{S D_0}{S-1} exp(-\beta_{AE} \sqrt{2/S}(r-r_0))
\ee
Hence, in fact we fit parameters $S, \beta_{AE}$ keeping $D_0$ and $r_0$ fixed.
The parameters in the angular part are identical with that of Tersoff.
The conversion of the obtained parameters to those used in the Tersoff
potential is straightforward \cite{GaN,SiC,pontifix}.
The required radial Tersoff parameters $A, B, \lambda_1$ and $\lambda_2$ can be
expressed using the Albe-Erhart parameters as follows:
\be
\lambda_1=\beta_{AE} \sqrt{2S}, \lambda_2=\beta_{AE} \sqrt{2/S}.
\ee
\beqn
A=D_0/(S-1)exp(\lambda_1r_0), && \\
B=SD_0/(S-1)exp(\lambda_2r_0),
\eeqn
This conversion must be done when e.g. the LAMMPS code is used.
Example file can be found in the released packages of LAMMPS.

 {\em The details of the fitting procedure:}
 We set $\Theta_0 \approx 80^{\circ}$ in order to get regular shape gr-humps. At $\Theta_0=90^{\circ}$
the shape of the bulges becomes irregular (e.g. elongated) at 300 K.
Also, in other cases (when various $D_0$ and $r_0$ values are set)  $\Theta_0=90^{\circ}$ leads to bulged-in regions
instead of humps and the overall gr-sheet becomes strained.
The cutoff parameter $R_c$ is kept at or below 3.0 $\hbox{\AA}$.
The increase of $R_c$ leads to improper surface features and to the
increase of hump height or in other cases to bulged-in regions (holes-like features). We find the maximal reasonable value
for $R_c \approx 3.0$ $\hbox{\AA}$.
Surprisingly the application of a long-range (dispersion-like) potential
with or without cutoff leads to serious overbinding of the humps.
This is what happens in particular with the Lennard-Jones potential,
which has a long range tail.
We managed to stabilize humps with a small corrugation level using
an additional Gauss-like potential within a tiny long-range region,
however in this case we get a double-well potential instead of a 
dispersion-like one. 

  The reduction of corrugation $\xi$ is one of the hardest challenges.
In general, the decrease of $\xi$ below $2$ $\hbox{\AA}$ is rather hard
and can only be done with the perceptible loss of other quality features.
In principle, the increase of $D_0$ (which makes C-Ru bond stronger) could reduce corrugation, however,
the adhesion energy is affected seriously by the magnitude of $D_0$.
On the other hand the decrease to $D_0 \approx 0.3$ eV results in
the increase of the hump height up to $3.0 \pm 0.3$ $\hbox{\AA}$.

  The diameter and partly the shape of the humps can be controlled partly by the ratio $R_c/D_c$.
Smaller the ratio larger the diameter of the bulges.
The controll of the shape of the protrusions is an even more difficult
task.
The transition between trigonal and hexagonal or even rounded hexagonal (spherical or dome-like) shapes can be seen under various
parameter sets. As mentioned before, irregular shapes with various
size can also been observed (or the mixture of them).
STM images, depending on the applied imaging technique mostly report trigonal or hexagonal domed shape humps with $\sim 1.5$ nm diameter \cite{gr-Ru_NM,gr-Ru,Parga,STM:Ru}.
Since it is often not easy to identify the exact shape of the moir$\acute{e}$ surface features (due to the insufficient resolution),
e.g. hexagonal often looks like a dome (and vice versa), or hexagons
can be distorted in such a way that looks more like trigonal protrusions,
it is important to study in detail the various possible morphologies
which appear as moir$\acute{e}$ superstructures during simulations.
It could also be usefull to develop different parameter sets which
provides different hump shapes.

 In general, 
the development of a reliable parameter set is required many trial-and-error
simulations (few hundreds) besides the optimization of the parameter set (using PONTIFIX).
In each step, after the manual change of the selected parameters ($D_0$, $r_0$, $R_c$, $D_c$)
the rest of the parameters (mostly the angular part) are optimized by PONTIFIX \cite{pontifix}.
With the new parameter set a new MD simulation is done at 0 K in order
to check its performance.
This is repeated many times until the best possible topography is obtained.
The main requirements are the following: the average corrugation is being below $2$ $\hbox{\AA}$ even at 300 K,
moreover, the 0 K structure should be stable at 300 K with
minor corrugation increase,
minimal C-Ru distance $d_{min} > 1.9$ $\hbox{\AA}$,
maximal C-Ru distance $d_{max} < 4.0$ $\hbox{\AA}$,
correct $12 \times 12$ unit supercell,
the adhesion energy is $0.15 < E_{adh} < 0.25$ eV/C,
no decorations occurs on the surface besides the regular shaped humps
(no further protrusions, vacancy islands, or holes).
These conditions are hard to be satisfied together. The last one is in particular
not fully satisfied, since we notoriously find small features like
thin (one atomic thick) line of protrusions (atomic chains) which did not appear in most of the
STM topographies available. However, these features can hardly be seen 
and the average line of protrusion height is less than $0.5$ $\hbox{\AA}$.
Within the Tersoff interface model we could not get rid of these features
completely which, however, did not affect seriously the otherwise nearly perfect
moire superstructures.
Sometimes, the maximal C-Ru distance $d_{max} > 4.0$ $\hbox{\AA}$ 
and can reach $d_{max} \approx 5.0$ $\hbox{\AA}$ or more.
This is due to the lack of long-range interactions which could lower
hump height.
The adhesion energy $E_{adh}$ and the
average corrugation are counter-related therefore it is especially
hard to find reasonable parameter values which keep both of them
within the required regime.

\begin{table}[t]
\caption[]
{
The fitted parameters used in the Tersoff expression fitted by
the PONTIFIX code using a small training set of weakly bound
alloys including only molecular dimer, NaCl (B1) and CsCl (B2) type unit cells.
Three sets of parameters are given for the irregular hexagonal,
trigonal and hexagonal domes.
}

\begin{ruledtabular}
\begin{tabular}{lccc}
C-Ru (Tersoff, pw)  & irreg. dome & trigonal & hexagonal     \\
\hline%

A (eV)    & 48.9256612   & 35.6806608 & 59.6418650 \\
B (eV)    & 7.34248258   & 10.0452586 & 18.8697839  \\
$\lambda_1$ & 2.60152968 & 3.75198031 & 3.5582675  \\
$\lambda_2$ & 0.76877847 & 0.53305183 & 0.56207129      \\
$\gamma$    & 0.04376490 & 0.10346188 & 0.13375989 \\
c           & 1.55708101 & 2.50168822 & 2.89790985  \\
d           & 0.33781624 & 0.22311075 & 0.19358384  \\
h           & -0.173648  & -0.17 & -0.17    \\
$R_c$ ($\hbox{\AA}$)         & 3.0 & 2.6 & 2.6   \\
$D_c$ ($\hbox{\AA}$)         & 0.2 & 0.3 & 0.6      \\
$\beta$ ($\hbox{\AA}^{-1}$)  & 1.0 & 1.0 & 1.0   \\
n                            & 1 & 1 & 1   \\
$\mu$                        & 0.0 & 0.0 & 0.0     \\
\hline
 C-C \cite{SiC}  & & &    \\
\hline
A (eV)    & 2019.8449 & &  \\
B (eV)    & 175.426651 & &   \\
$\lambda_1$ & 4.18426232 & &    \\
$\lambda_2$ & 1.93090093 & &   \\
$\gamma$                     & 0.11233 & &  \\
c                            & 181.91 & &    \\
d                            & 6.28433 & &   \\
h                            & 0.5556  & &   \\
$R_c$ ($\hbox{\AA}$)         & 2.0 & &   \\
$D_c$ ($\hbox{\AA}$)         & 0.15 & &      \\
$\beta$ ($\hbox{\AA}^{-1}$)  & 1.0 & &  \\
n                          & 1 & &   \\
$\mu$                      & 0.0 & &    \\
\end{tabular}
\end{ruledtabular}
\footnotetext[1]{
pw: present work,
The parameters have been fitted to a small data base 
which includes dimer, NaCl (B1) and CsCl (B2) type unit cells
with fictitious bond distances ($d \approx 1.7-2.0$, $\hbox{\AA}$).
For fitting the PONTIFIX \cite{pontifix} code has been utilized.
The C-C Abell-Tersoff potential is also given (Erhart-Albe parameterization \cite{SiC}
of the Albe-potential, parameters are converted into the form suitable for
Tersoff formalism using Eqs. (10)-(14)).
}
\label{T1}
\end{table}

\begin{figure*}[hbtp]
\label{F1}
\begin{center}
\includegraphics*[height=6cm,width=9cm,angle=0.]{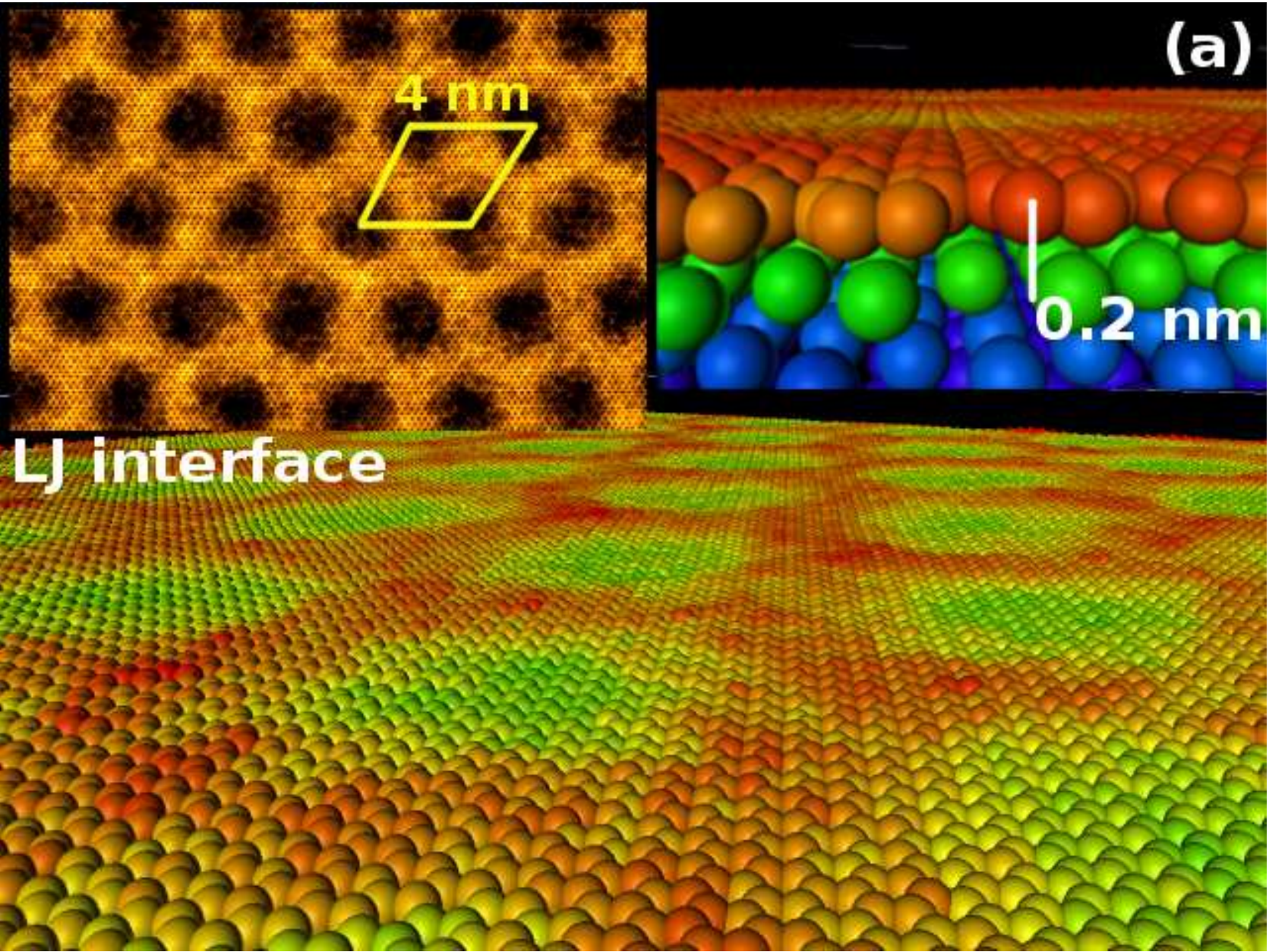}
\includegraphics*[height=6cm,width=8cm,angle=0.]{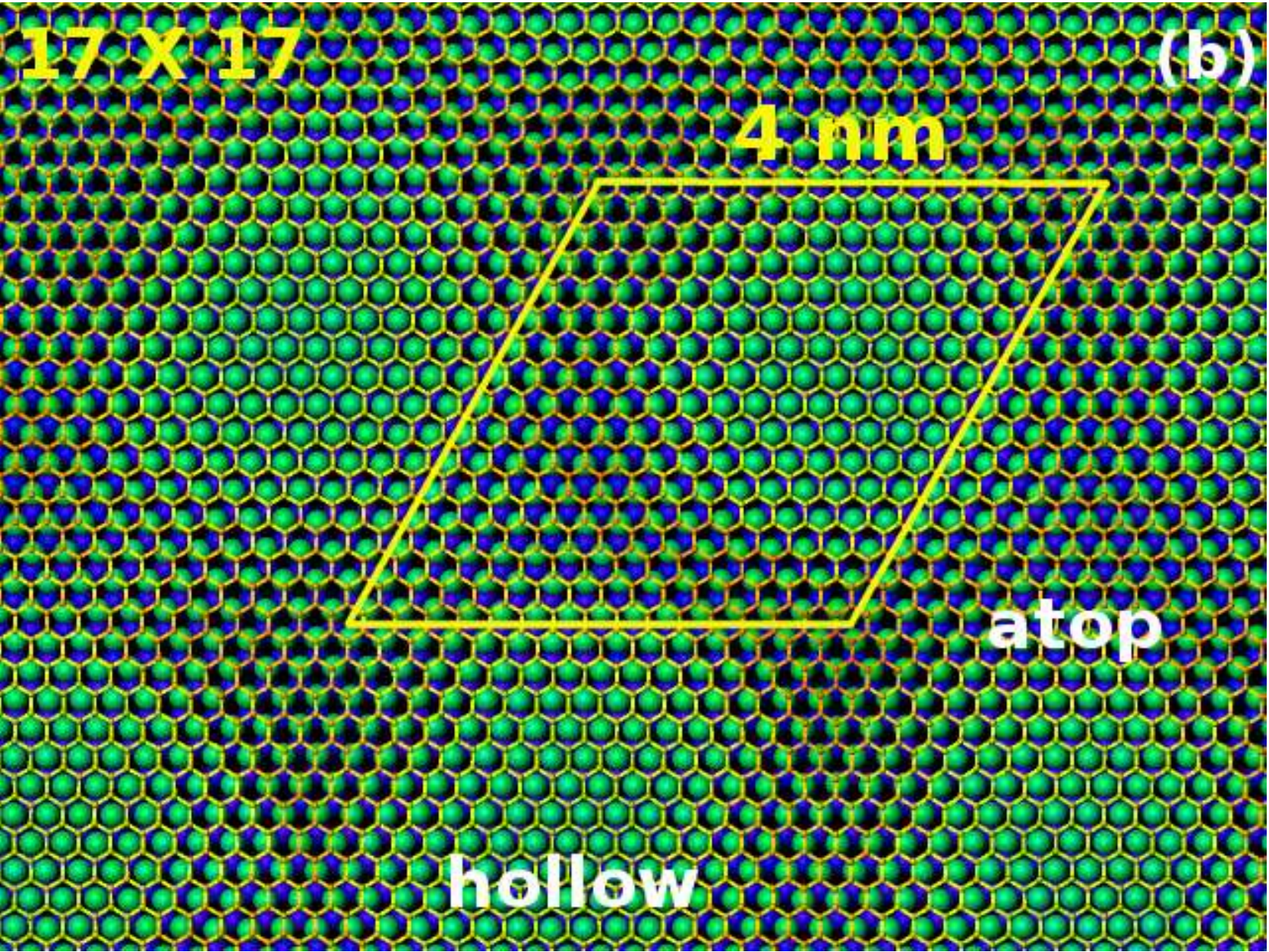}
\caption[]{
 The results of CMD simulations with Lennard-Jones potential
($\epsilon=0.1$, $\sigma=2.0$) at the interface.
(a) The size of the supercell is shown ($17 \times 17$ unit cell,  $4$ nm)
together with the cross-section of the interface.
The top-view (black regions are the bumps) and the landscape are also depicted (red regions
correspond to the protrusions).
The adhesion energy is -0.27 eV/C.
(b) The rhomboid supercell is shown with various registry sites.
Note that the hollow sites correspond to the bumps (valleys)
and the atop positions to the humps (protrusions).
This is an incorrect registry.
}
\end{center}
\end{figure*}

\subsection{Simulation rules}

 Before the MD simulations geometry optimization has been applied
using the fire algorithm together with the box/relax option
which allows 
the simulation box size and shape to vary during the iterations of the minimizer so that the final configuration will be both an energy minimum for the potential energy of the atoms, and the system pressure tensor will be close to the specified external tensor.

 Isobaric-isothermal (NPT ensemble) simulations (with Nose-Hoover thermostat and prestostat) were carried out at 300 K, vacuum regions were inserted
between the slabs of the gr-substrate system to ensure
the periodic conditions not only in lateral directions (x,y) but also in perpendicular direction to the gr sheet (z).
The variable time step algorithm has been exploited.
Few bottom layers of the substrate have been fixed in order to
rule out the rotation or the translation of the simulation cell.
The code OVITO \cite{ovito}
has been utilized for displaying atomic and nanoscale structures \cite{Sule_2011}.
 The system sizes of $5 \times 5$ nm$^2$ up to $100 \times 100$ nm$^2$
have been simulated under parallel (mpi) environment.

\begin{figure*}[hbtp]
\begin{center}
\includegraphics*[height=5cm,width=5.5cm,angle=0.]{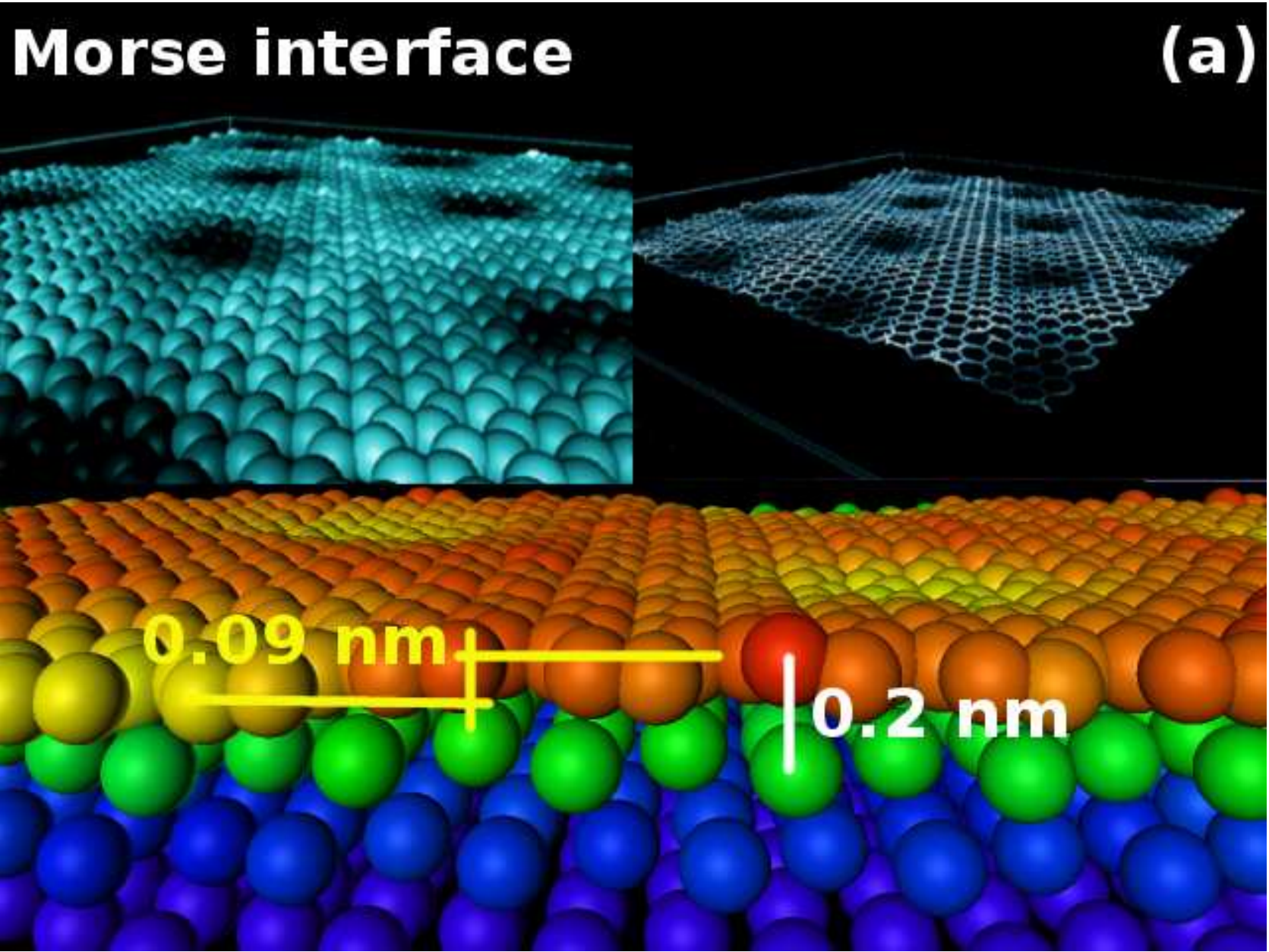}
\includegraphics*[height=5cm,width=6.4cm,angle=0.]{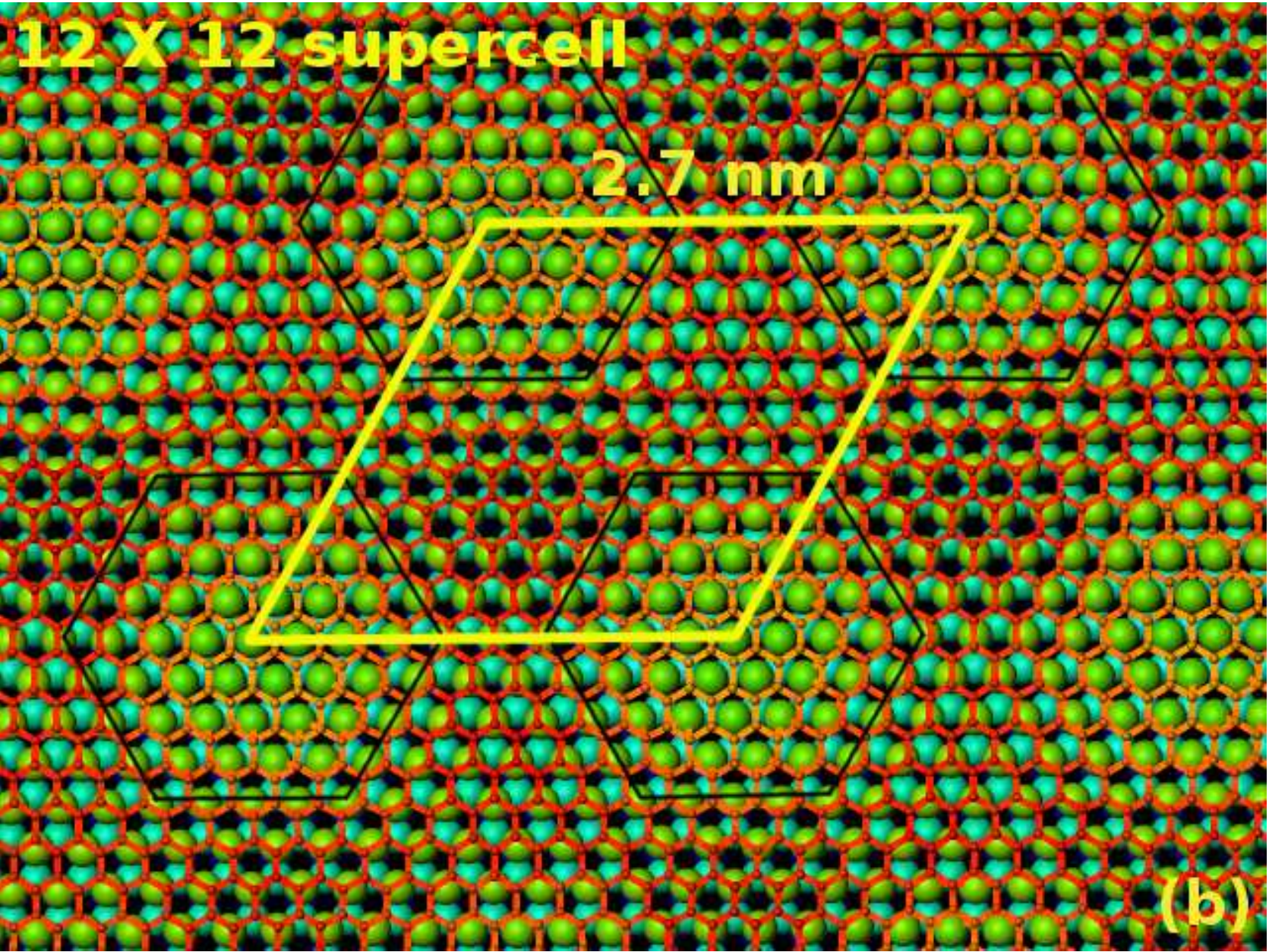}
\includegraphics*[height=5cm,width=5.5cm,angle=0.]{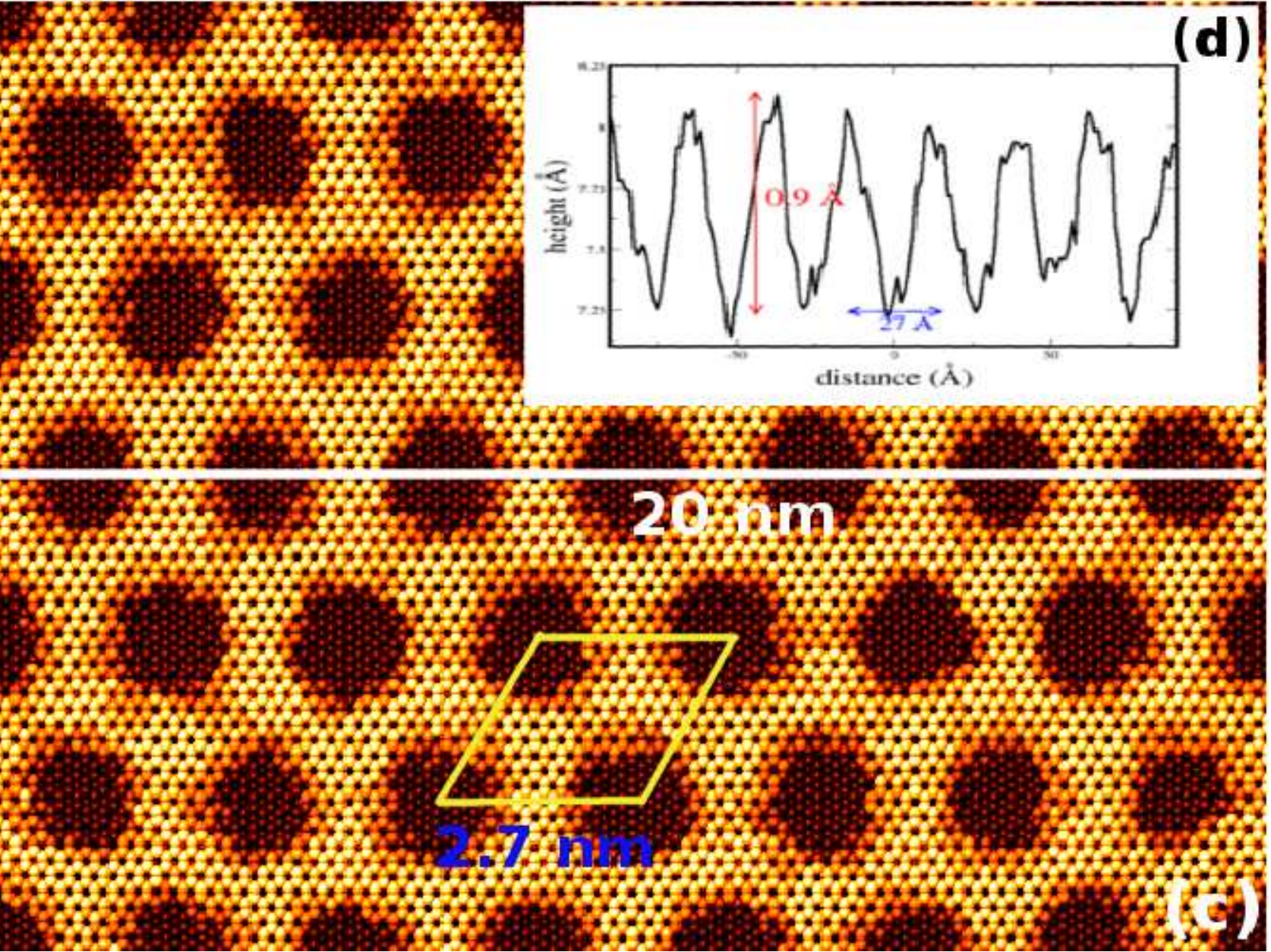}
\caption[]{
Simulation results obtained by the simple Morse potential.
(a): Small system with 965 Carbon atoms.
The top view with the registries (a) and the
cross-sectional view of the interface (b) together
with the inserted landscape views. 
The corrugation is rather low ($0.8 \pm 0.1$ $\hbox{\AA}$)
and the adhesion energy $E_{adh} \approx -0.17$ eV/C.
(b)-(d): Simulation results obtained 
for larger systems.
Moir$\acute{e}$ superstructure and the supercell.
The subregions with hollow sites (hollow-bumps) are marked by hexagons (b). The center of these
hexagons correspond to the rhomboid superstructure.
The regions marked by hexagons are the bumps and those in between
them are the humps. The latter one includes incorrectly fcc and hcp atop sites
which should be bulged-in regions. 
(c): A larger area is shown with hollow-bumps (dark spots).
The corresponding height profile is also shown (d).
}
\label{xl}
\end{center}
\label{F2}
\end{figure*}

\begin{figure*}[hbtp]
\begin{center}
\includegraphics*[height=5cm,width=6.5cm,angle=0.]{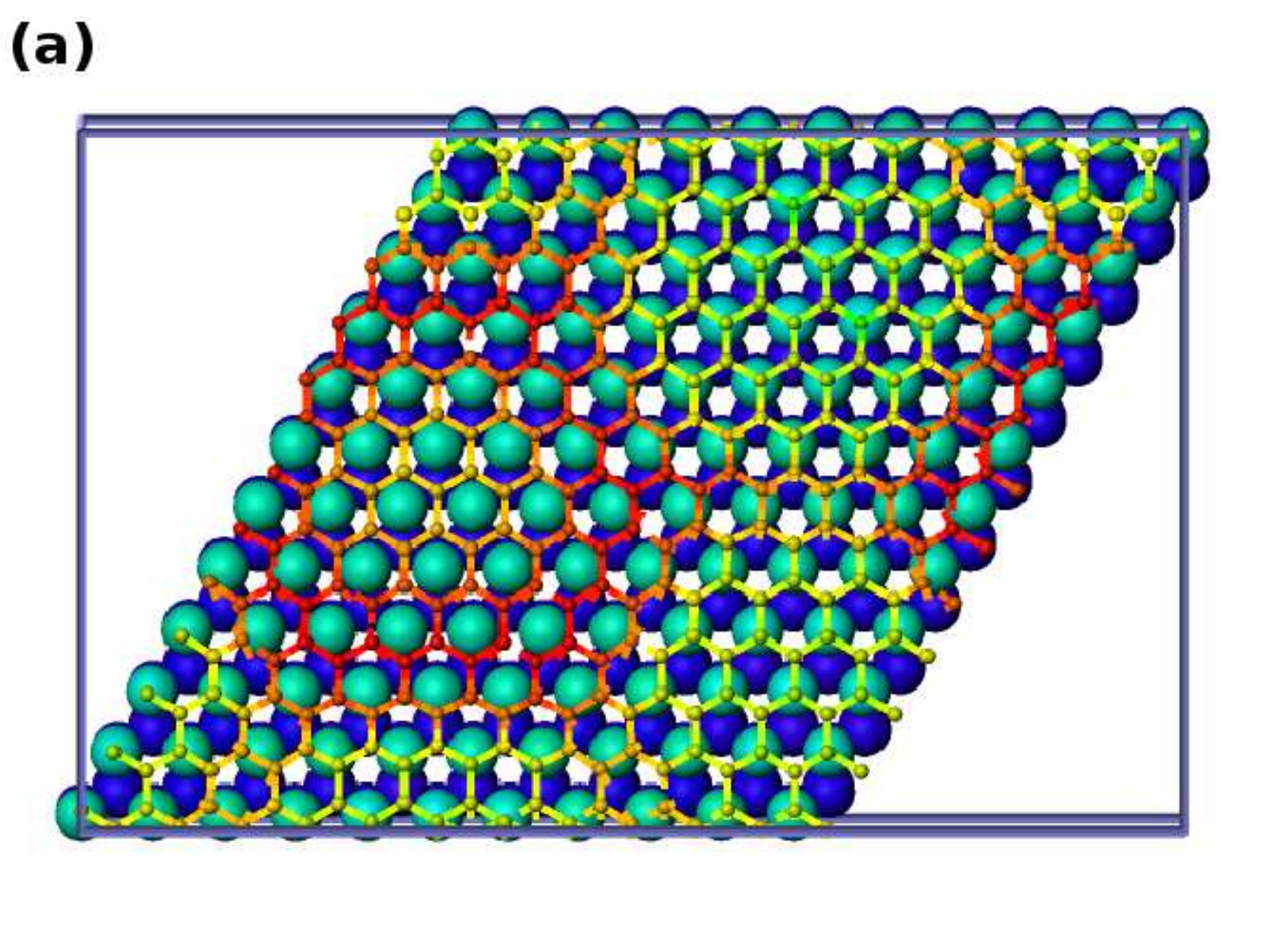}
\includegraphics*[height=5cm,width=6.5cm,angle=0.]{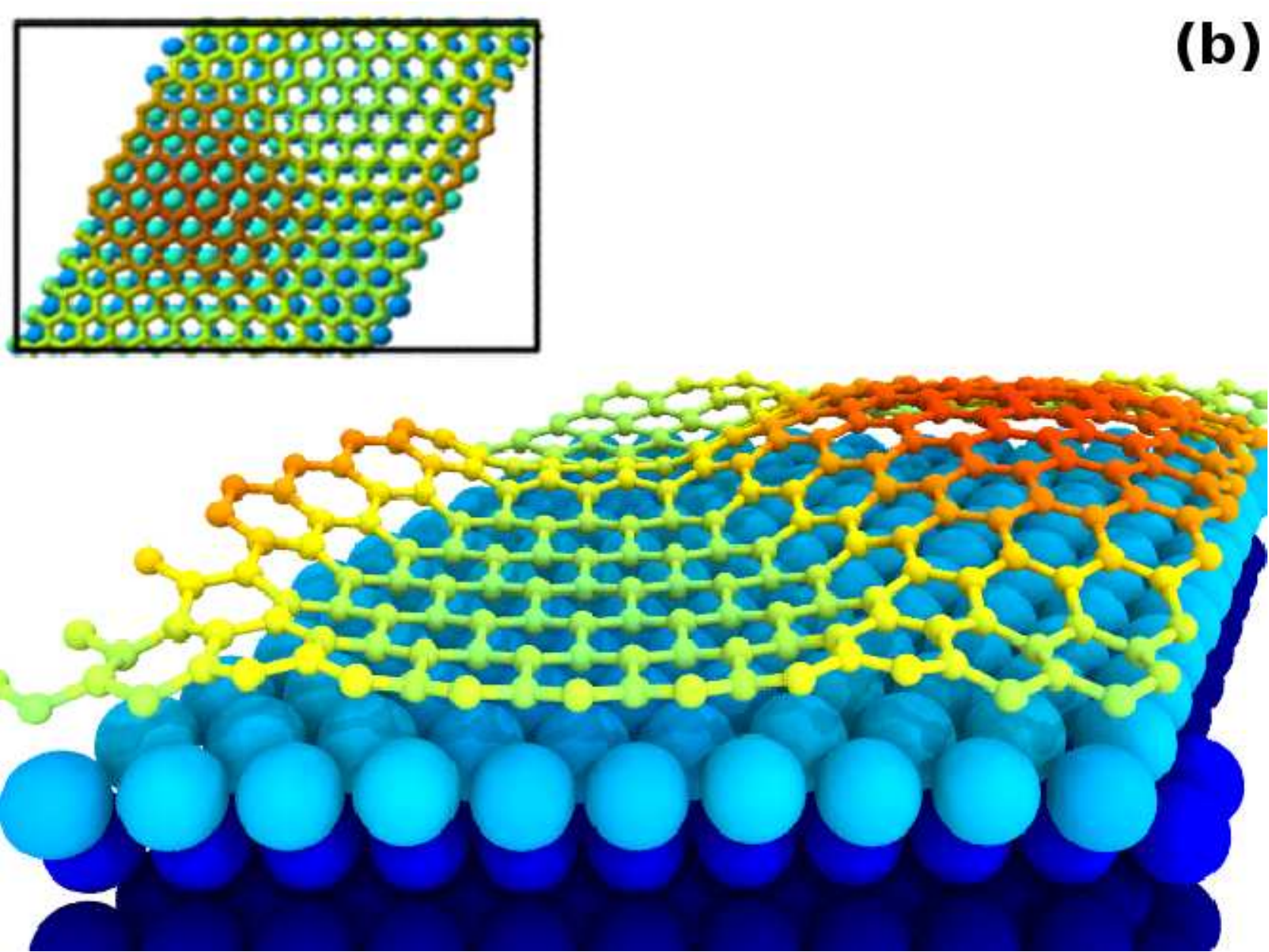}
\includegraphics*[height=5cm,width=6.5cm,angle=0.]{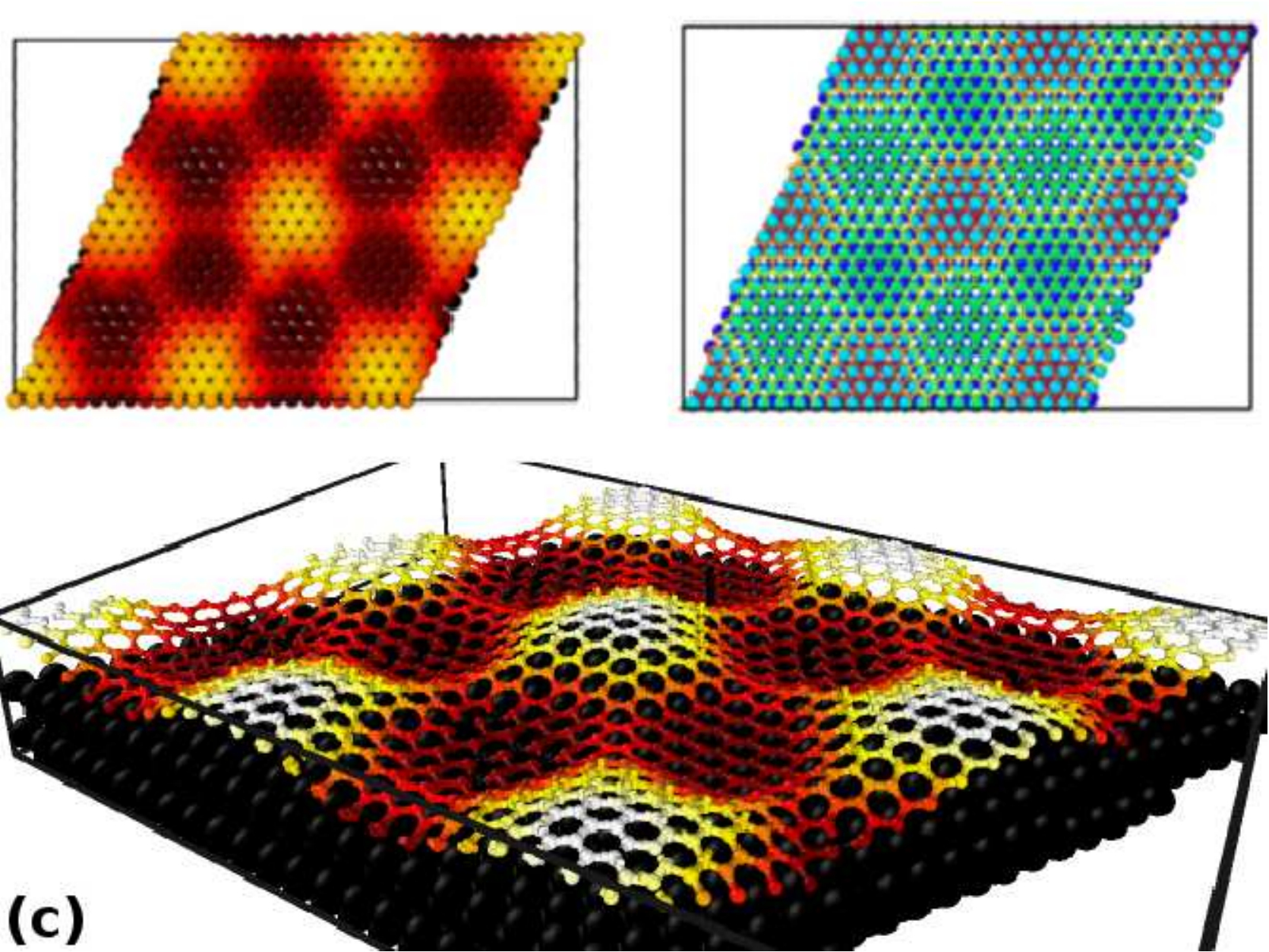}
\includegraphics*[height=5cm,width=6.5cm,angle=0.]{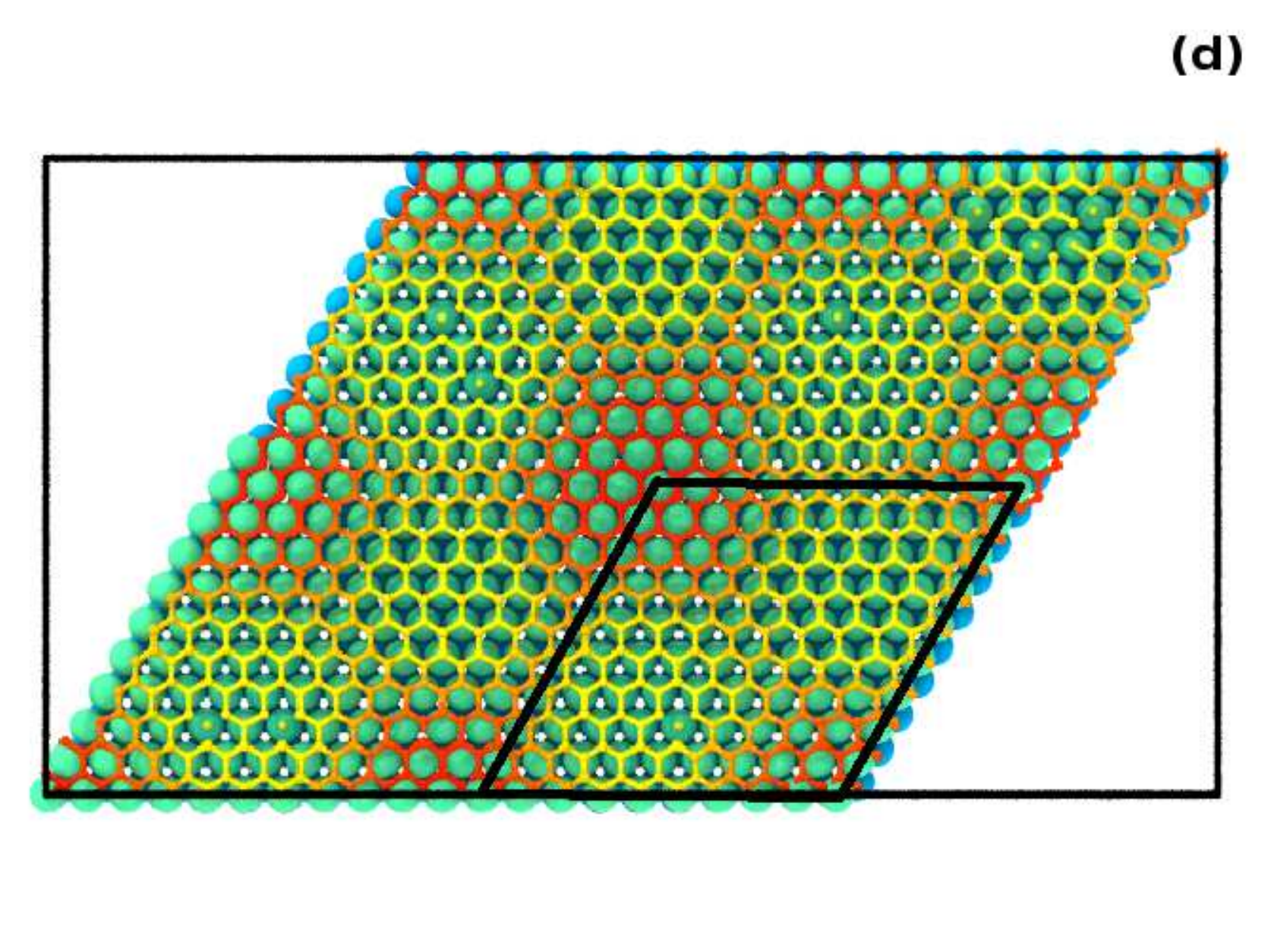}
\caption[]{
Simulation results obtained by the parameterized Tersoff-only potential.
(a) The minimal rhomboid supercell of the system (302 Carbon atoms, $12 \times 12$ gr unit cell
on $11 \times 11$ Ru unit cell. The perspective view is also given
(b) together with the top view of the wire model.
(c) A larger superstructure (948 Carbon atoms) which includes 4 equivalent subunit which correspond
to the minimal supercell.
(d) The top view of the 23on25 system with various registries.
The minimal supercell is also shown as a subunit marked by a rhombus.
Hollow sites (hollow humps, red wires) can be seen e.g. in the middle of the
superstructure. On-top sites (yellow or lighter regions)
correspond to the flatter areas (on-top bumps).
bulged-in (bumps) regimes.
This figure already shows the correct registry characteristics in accordance
with DFT calculations.
}
\end{center}
\label{F3}
\end{figure*}

\begin{table}[t]
\raggedright
\caption[]
{
The peak-to-peak corrugation ($\hbox{\AA}$) obtained for gr/Ru(0001)
by various experimental or DFT methods and compared with
CMD results obtained for dome-like superstructures.
}
\begin{ruledtabular}
\begin{tabular}{lcc}
 method (EXP, DFT)  & gr & Ru (topmost) \\
\hline
SXRD-1 & $1.5$ & $0.2$ \\
SXRD-2 & $0.82 \pm 0.15$ & $0.19 \pm 0.02$ \\
LEED & $1.5$ & $0.23$  \\
HAS & $0.15-0.4$ & \\
STM & $0.5-1.1$ & \\
DFT-PBE-1 & $1.75$ & \\
DFT-PBE-2 & $1.6$ & $0.05$ \\
DFT-vdW1 & $1.17$ & \\
DFT-vdW2 & $1.2$ & $0.045$  \\
\hline \hline
present work &   &  \\
\hline
CMD (Morse) & $1.1 \pm 0.1$  & $0.2 \pm 0.05$ \\
CMD (LJ) & $0.8 \pm 0.2$  & $0.4 \pm 0.1$   \\
CMD (Tersoff) & $1.9 \pm 0.1$  & $0.5 \pm 0.1$  \\
T,min (Tersoff)  & $1.9 \pm 0.1$  & $0.7 \pm 0.2$  \\
DFT/vdW-DF2/CMD & $2.0 \pm 0.2$  & $0.4 \pm 0.1$  \\
\end{tabular}
\footnotetext[1]{
SXRD-1: surface X-ray diffraction \cite{SXRD:25X25}, 
SXRD-2: \cite{Xray:Ru}, LEED: Low-energy electron diffraction \cite{LEED:Ru},
HAS: Helium Atom Scattering \cite{Parga}, 
STM: Scanning Tunelling Microscope \cite{STM:Ru}, 
DFT-PBE-1 \cite{DFT:Ru_corrug},
DFT-PBE-2 \cite{DFT:Ru_Wang},
DFT-vdW1 : Density Functional Theory (PBE-Grimme, D2) see in ref. \cite{DFT:Ru-Hutter},
DFT-vdW2 (PBE-Grimme, D2) see in ref. \cite{DFT:Ru-Stradi},
CMD: classical molecular dynamics (present work) obtained
using the Morse and LJ potentials.
{\em T,min}: geometry optimization and energy minimization only (molecular mechanics, no MD) using the steepest-distant
method for finding the energy minimum of the system together
with anisotropic box relaxation (Tersoff-only). This method provides results
which can be compared directly with {\em ab initio} DFT geometry optimization
results.
DFT/DF2/CMD: The DFT/revPBE-DF2 vdW-functional \cite{revPBE,Dion} is used
as imlemented in the trunk version of SIESTA (LMKLL) \cite{SIESTA2}
including 748 atoms (302 Carbon atoms) with a minimal supercell.
}
\label{T2}
\end{ruledtabular}
\end{table}

\subsection{{\em Ab initio} DFT calculations}

 First principles DFT calculations have also been carried out for
comparing structural and energetical properties with MD simulations.
Although we find a large number of published DFT results in the literature
for gr/Ru(0001) (see e.g. refs. \cite{DFT:Ru-Stradi,DFT:Ru-Hutter,Peng}),
however, in one hand we tried to reproduce these results 
and on the other one we utilize a somewhat different conditions.
In particular, geometry optimization has been carried out in two ways:
at first we start from a flat system (as usual) to reproduce the available DFT
results published recently \cite{Peng}.
We do not impose periodic boundary conditions (PBC) to avoid possible artifacts
caused by PBC at the edge of the simulation cell \cite{Peng}.

 At second, 
the DFT structural relaxations are initialized from
our "preoptimized" CMD structure in order
to see how the CMD structure relaxes.
The reason is twofold: we would like to calculate the adhesion energy
of the CMD structure by DFT (single point calculation only) and
also one can see how close the CMD structure is to the DFT optimal one
(using conjugate gradient geometry optimization).

 For this purpose we used 
the SIESTA code \cite{SIESTA,SIESTA2} which utilizes atomic centered numerical basis set.
The SIESTA code and the implemented Van der Waals functionals (denoted as DF1 and DF2,
DRSLL and LMKLL in the code \cite{SIESTA2} 
have been tested in many articles for gr (see e.g recent refs. \cite{sies1,sies2}).
We have
used Troullier Martin, norm conserving, relativistic pseudopotentials in fully separable
Kleinman and Bylander form for both carbon and Ru.
Throughout the geometry optimization, numerical atomic orbitals
first single-$\zeta$ (SZ) and then double-$\zeta$ polarization (DZP) basis sets were
used.
In particular, 16 valence electrons are considered for Ru atoms
and 4 for C atoms.
Only $\Gamma$ point is used for the k-point grid in the SCF cycle.
The real space grid used to calculate the Hartree, exchange and correlation
contribution to the total energy and Hamiltonian was 200 Ry (Meshcutoff).
 The gradient-corrected Exchange and correlation are calculated
by the revPBE/DF2 van der Waals functional \cite{Dion}.
We also employed the local density approximation (LDA) and the
gradient corrected
variant PBE \cite{PBE}.

 The system consists of 300 Carbon and 478 Ru atoms (4 layers Ru).  
The bottom Ru layer is fixed.
The minimal rhomboid supercell similar to that shown in Fig. 3(a) has been used as
an input structure.
Conjugate gradient geometry optimization has been employed in each cases
until the rms forces went below 0.001  eV/$\hbox{\AA}$ and interatomic forces were smaller 
than 0.05 eV/$\hbox{\AA}$.

\section{Results and Discussion}

\subsection{Results for the Lennard-Jones and Morse pairwise potentials}

 First results will be shown briefly obtained by the simple
Lennard-Jones potential (see Fig. 1(a)-(b)).
The LJ potential not only provides wrong registry but also
the supercell size is exaggerated: $\sim 4$ nm instead of $2.7$ nm.
The shape of the rhomboid supercells are somewhat distorted and the topography becomes
disordered at lower adhesion energy ($E_{adh} < -0.25$ eV/C).
The corrugation is too low ($\xi \approx 0.08$ nm).
Hence we find LJ potential is inappropriate for the description of
gr Moir$\acute{e}$ superstructures.

 Not only LJ, but other simple pair-wise potentials, such as Morse
 potential
is unable to reproduce Moir$\acute{e}$ superstructures.
After visual inspection of Figs. 2 one can conclude that
the Morse potential provides also incorrect topography: it favors energetically
hollow site vs. on-top configurations.
This is again in contrast with DFT results which show the contrary results \cite{Batzill}.
CMD-Morse overbinds hollow sites vs. DFT which binds more strongly fcc and hcp sites.
CMD gives the adsorption energy of -0.222 eV/C for a purely hollow registry, -0.082 eV/C for atop configurations and -0.172 eV/C for a mixed system (hollow and other sites).
The only advantage of the Morse function is that the obtained
supercell size is correctly $2.7$ nm.

 The peak-to-peak corrugation ($\xi_{corr}$) is averaged for small subregions which
includes the $12 \times 12$ superstructure.
The sampling of these regions was taken on large systems with a lateral size
up to $100 \times 100$ nm$^2$.
We find that the overall average surface corrugation remains 
in the range of $\xi \approx 1.1 \pm 0.1$ $\hbox{\AA}$ 
when the adhesion energy of the gr-layer is
$E_{adh} \approx -0.17$ eV/C atom.

  The corresponding literature data on corrugation obtained by various experimental and DFT methods
are summarized in Table ~\ref{T2}.
In spite of the great efforts to obtain the structural corrugation
of the moir$\acute{e}$ observed in gr/Ru, the magnitude of
the corrugation is still a subject of controversy.
In general, the data
scatters within the wide range of $0.15-1.75$ $\hbox{\AA}$.
The lower range is provided mostly by
HAS \cite{Parga} and SXRD \cite{SXRD:25X25} experiments and partly by STM \cite{STM:Ru}.
The DFT/LDA results of Peng {\em et al.} support low corrugation of $0.2$
$\hbox{\AA}$ (essentialy flat gr) where they used a large quasi-periodic system
claiming that in other DFT calculations the imposed
periodic conditions lead to artificially large corrugation \cite{Peng}.
Other experiments, such as SXRD \cite{Xray:Ru}, STM \cite{STM:Ru}, LEEM \cite{LEED:Ru}  and DFT methods \cite{DFT:Ru_corrug,DFT:Ru_Wang},
provided larger corrugation values in the range of $0.5-1.75$ $\hbox{\AA}$.
  STM values are strongly influenced by the applied bias voltage \cite{STM:Ru}.
Nevertheless, the obtained pattern is clearly observable and the moir$\acute{e}$
supercell size corresponds to the widely accepted $12 \times 12 (gr)/11 \times 11 (Ru)$ \cite{rev,Batzill} one.
It is important to note that both nearly flat ($\xi < 0.5$ $\hbox{\AA}$) and corrugated
gr is observable under different but not yet clearly understood conditions.
Partly it is our motivation to develop a new force-field model which could
explain this controversy.
Using large scale atomistic simulations it could be possible
to clarify in more detail the relative energetics of various corrugation
level superstructures.

\begin{figure*}[hbtp]
\begin{center}
\includegraphics*[height=5cm,width=5.5cm,angle=0.]{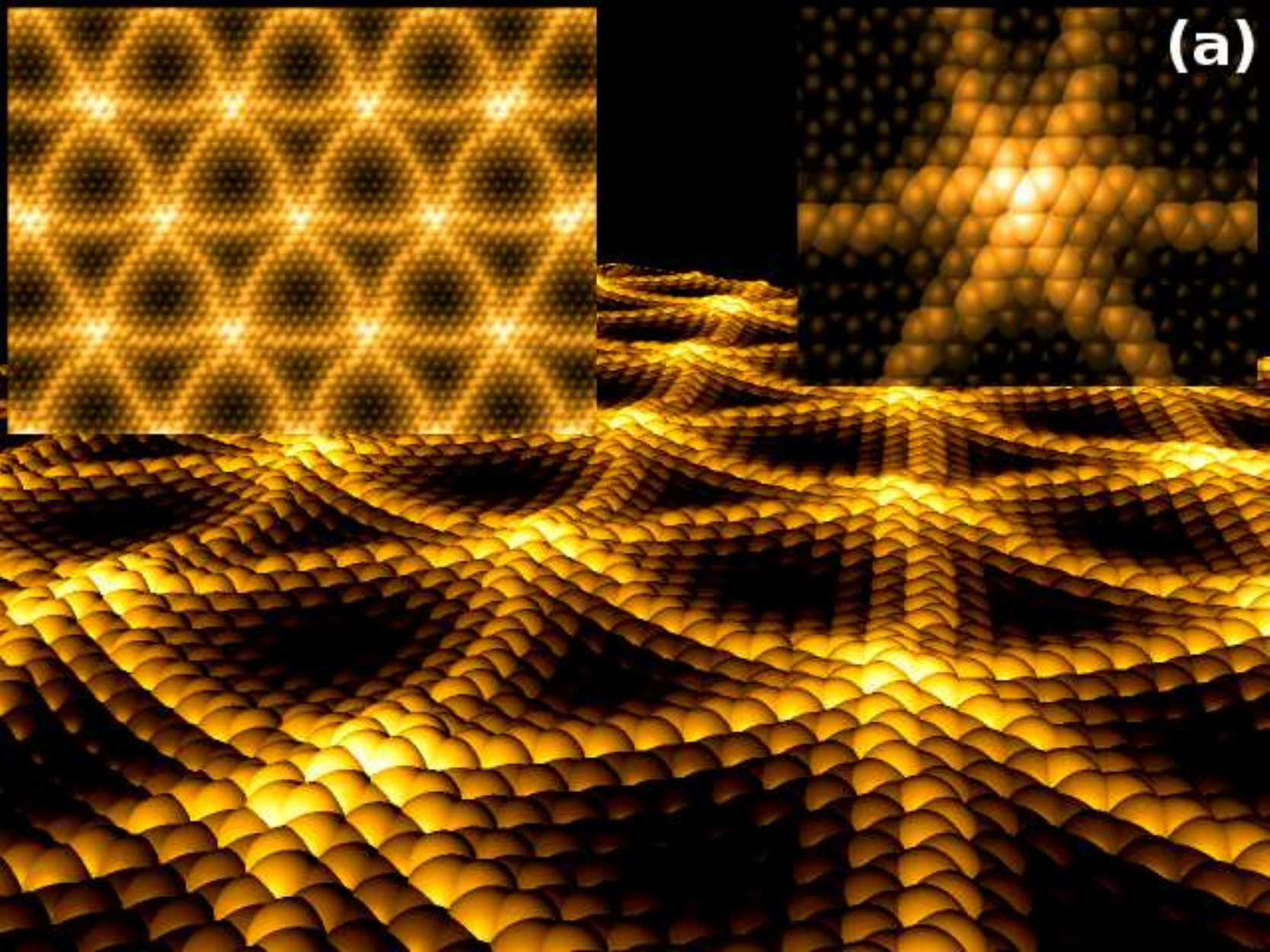}
\includegraphics*[height=5cm,width=5.5cm,angle=0.]{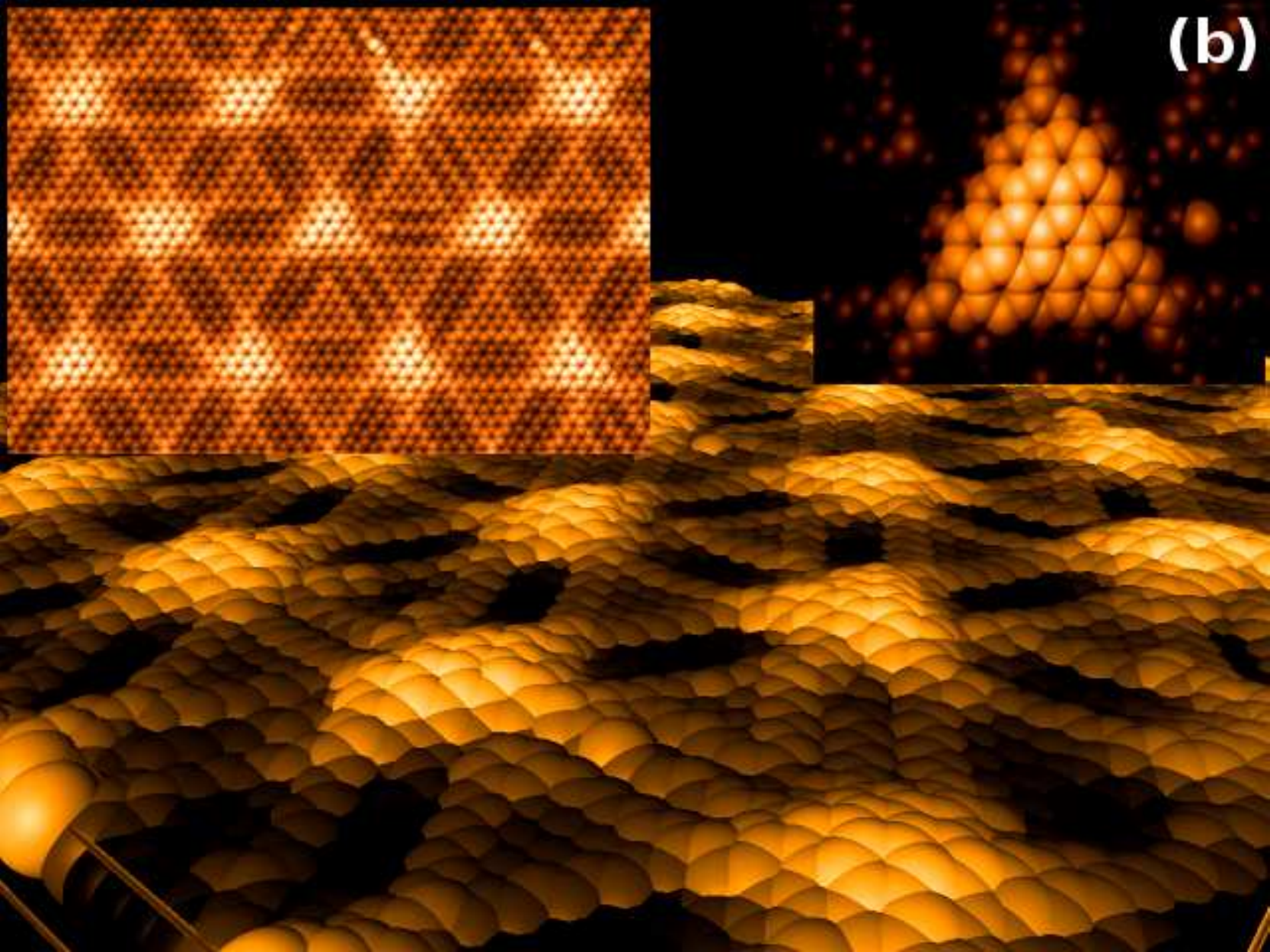}
\includegraphics*[height=5cm,width=6.4cm,angle=0.]{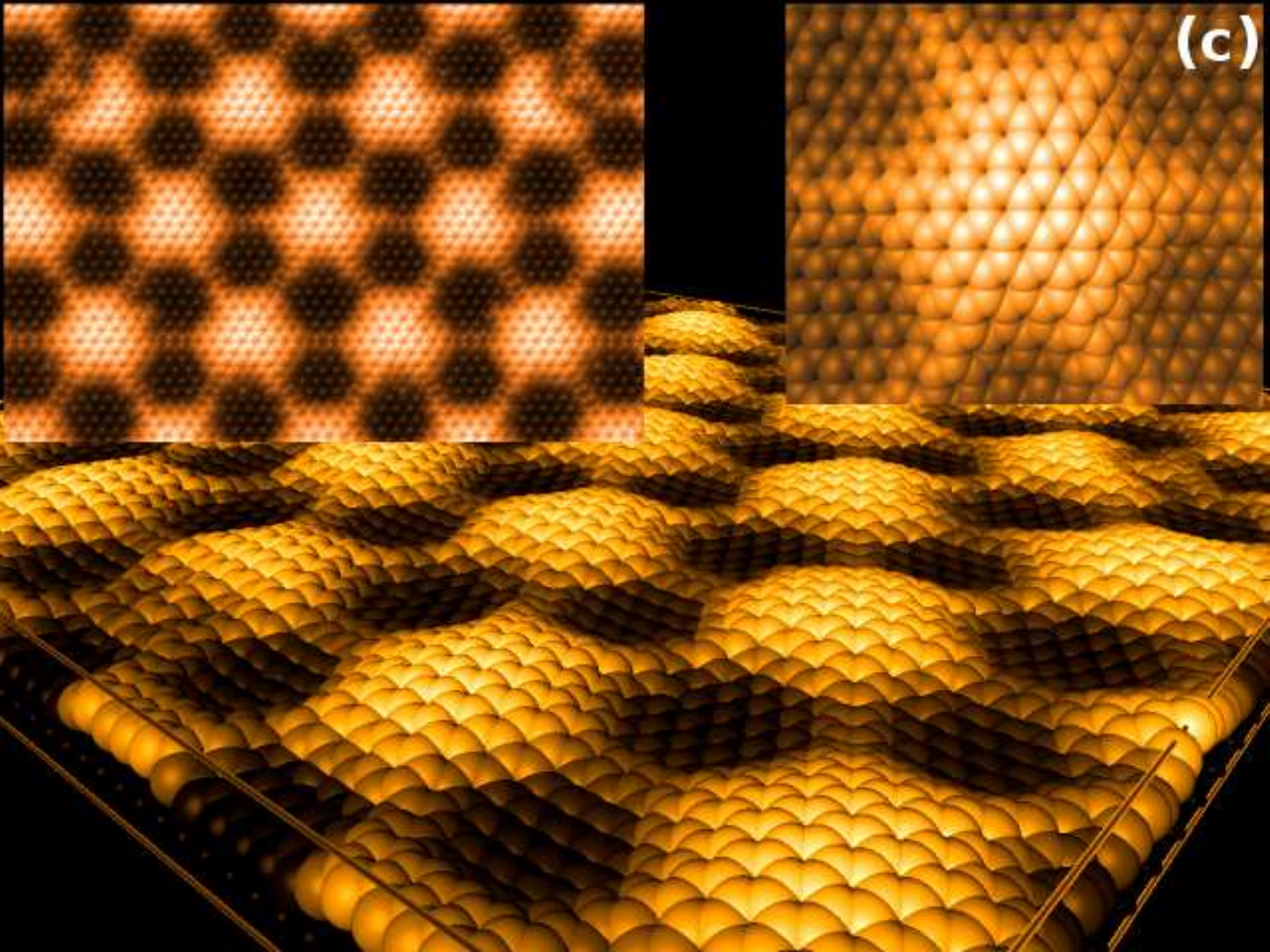}
\caption[]{
The landscape view of various morphological forms:
(a) irregular dome-like hexagonal protrusion with 6 facets, (b) tetrahedron (trigonal),
(c) hexagonal dome (6 facets).
The characteristic hump shapes found during the extensive search
for the optimal interfacial Tersoff parameter set are also shown
for each panel on the right.
The top view of the simulated gr-surface can be seen on the left
on each panel.
These polyhedral features perform nearly equally, non of them
can be taken as the most stable form.
}
\end{center}
\label{F4}
\end{figure*}
\begin{table*}[tb]
\raggedright
\caption[]
{
The summary of various properties obtained for gr/Ru(0001)
by classical molecular dynamics simulations using the fitted Tersoff
potential for the interface.
The properties of moire superstructures with different shapes
(irregular hexagonal dome, trigonal (tetrahedral) and regular hexagonal dome) are compared.
}
\begin{ruledtabular}
\begin{tabular}{lcccccc|ccc}
  pw & $\xi$ ($\hbox{\AA}$)  & $\xi_{Ru}$ ($\hbox{\AA}$) & $d_{min}$ ($\hbox{\AA}$) & $d_{max}$ ($\hbox{\AA}$) & $a_{gr}$ ($\hbox{\AA}$) & $a_{lm}$ ($\%$) & $E_{adh}$ (eV/C) & $E_{str}$ (eV/C) & $E_{gr}$ (eV/C) \\
\hline
  &  &  &  &  &  &  &  &  &   \\
 {\bf irregular dome} &  &  &  &  &  &  &  &  &   \\
  &  &  &  &  &  &  &  &  &   \\
 T,min & $2.2 \pm 0.2$ & $0.7 \pm 0.2$ & 1.9 & 4.1 & 2.54 & 5.38 & -0.088 & 0.32 & -6.9438  \\
 T,md (0 K) & $1.8 \pm 0.1$ & $0.5 \pm 0.2$ & 1.9 & 4.1 & 2.54 & 5.34 & -0.101 & 0.25 & -7.012  \\
 T,md (300 K) & $1.9 \pm 0.1$ & $0.5 \pm 0.2$ & 2.5 & 5.0 & 2.53 & 6.00 & -0.17 & 0.03 & -7.233  \\
  &  &  &  &  &  &  &  &  &   \\
 {\bf trigonal} &  &  &  &  &  &  &  &  &   \\
  &  &  &  &  &  &  &  &  &   \\
 T,md (0 K) & $2.3 \pm 0.2$ & $0.5 \pm 0.2$ & 2.2 & 4.5 & 2.55 & 5.78 & -0.301 &  0.28 & -6.984 \\
 T,md (300 K) & $2.4 \pm 0.2$ & $0.5 \pm 0.2$ & 2.2 & 4.9 & 2.57 & 5.16 & -0.301 &  0.058 & -7.208 \\
  &  &  &  &  &  &  &  &  &   \\
 {\bf hexagonal dome} &  &  &  &  &  &  &  &  &   \\
  &  &  &  &  &  &  &  &  &   \\
 T,md (0 K) & $2.4 \pm 0.2$ & $0.08 \pm 0.02$ & 2.0 & 5.9 & 2.55 & 5.28 & -0.366 &  0.02 & -7.241 \\
 T,md (300 K) & $2.5 \pm 0.2$ & $0.08 \pm 0.02$ & 1.9 & 6.2 & 2.56 & 5.20 & -0.348 &  0.03 & -7.230 \\
  &  &  &  &  &  &  &  &  &   \\
\hline
  &  &  &  &  &  &  &  &  &   \\
 DFT & 1.1-1.7$^b$ & 0.05$^b$ & 2.0$^{b,c}$ & 3.8$^{b,c}$ & 2.51-2.54$^c$ & 
 6-7$^c$ & -0.20$^{b,c}$ & 0.104$^{b,c}$ & n/a   \\
  &  &  &  &  &  &  &  &  &   \\
  {\bf DFT / present work:} &  &  &  &  &  &  &  &  &   \\
  &  &  &  &  &  &  &  &  &   \\
 DFT/vdW-DF2 (corr) & $2.15 \pm 0.2$ & $0.5 \pm 0.1$ & 2.05 & 4.45 &  
  2.56 & 6.3 & -0.125 & 0.06 & n/a   \\
 DFT/vdW-DF2 (flat) & $0.4 \pm 0.1$ & $0.1 \pm 0.02$ & 2.34 & 2.55 & 2.56 &  
  6.3 & -0.08 & 0.08 & n/a   \\
\end{tabular}
\footnotetext*[1]{
pw denotes present work,
$\xi$ and $\xi_{Ru}$ are the average corrugation for gr and the topmost
Ru(0001) layer ($\hbox{\AA}$).
$d_{min}$ and $d_{max}$ are the minimal and maximal inter-layer distances ($\hbox{\AA}$) at the interface.
$a_{gr}$, $a_{lm}$ are the lattice constant of gr ($\hbox{\AA}$) and
the lattice mismatch ($\%$) after simulations ($a_{lm}=100 (a_{s}-a_{gr})/a_{gr}$).
T, min: Tersoff-only results, geometry optimization only.
T,md: Tersoff-only results with CMD at 0 and 300 K.
TG,lr: results obtained by the Tersoff+long-range method (Tersoff-Gauss).
The long-range parameters are the following:
DFT results are also given for comparison \cite{DFT:Ru-Hutter,Batzill,DFT:Ru_Wang}.
DFT (pw): PBE \cite{PBE} and revPBE/vdW-DF2 \cite{Dion} results as obtained by the author. See further details
in the caption of Table II.
All quantities are given per Carbon atom, except the
atomic cohesive energy of the substrate.
The adhesion energy $E_{adh}=E_{tot}-E_{no12}$, where $E_{tot}$ is the potential energy/C
after md simulation. $E_{no12}$ can be calculated using the final
geometry of md simulation with heteronuclear interactions
switched off. Therefore, $E_{adh}$ contains only contributions from
interfacial interactions.
$E_{str,gr}$, the strain energy of the corrugated gr-sheet
include terms comming from stretching and corrugation (bending and torsional strain)
($E_{str,gr}=E_{gr}-E_{gr,flat}$),
where $E_{gr}$ and $E_{gr,flat}=-7.2662$ eV/C are the cohesive energy of C atoms
in the corrugated and in the relaxed flat (reference) gr sheet.
Using the SIESTA code (DFT/revPBE-DF2) the strain energy has also been
estimated at the ab initio level of theory for the hexagonal dome system.
{\em flat} system: DFT GO is also done for the flat gr/Ru(0001) system.
The geometry optimization is started from the CMD structure and has been continued
until the rms gradient went below 0.005 a.u.
\\
$^b$ from ref. \cite{DFT:Ru-Hutter},
$^c$ from ref. \cite{DFT:Ru_corrug}.
}
\label{T3}
\end{ruledtabular}
\end{table*}

\begin{figure*}[hbtp]
\begin{center}
\includegraphics*[height=8cm,width=10cm,angle=0.]{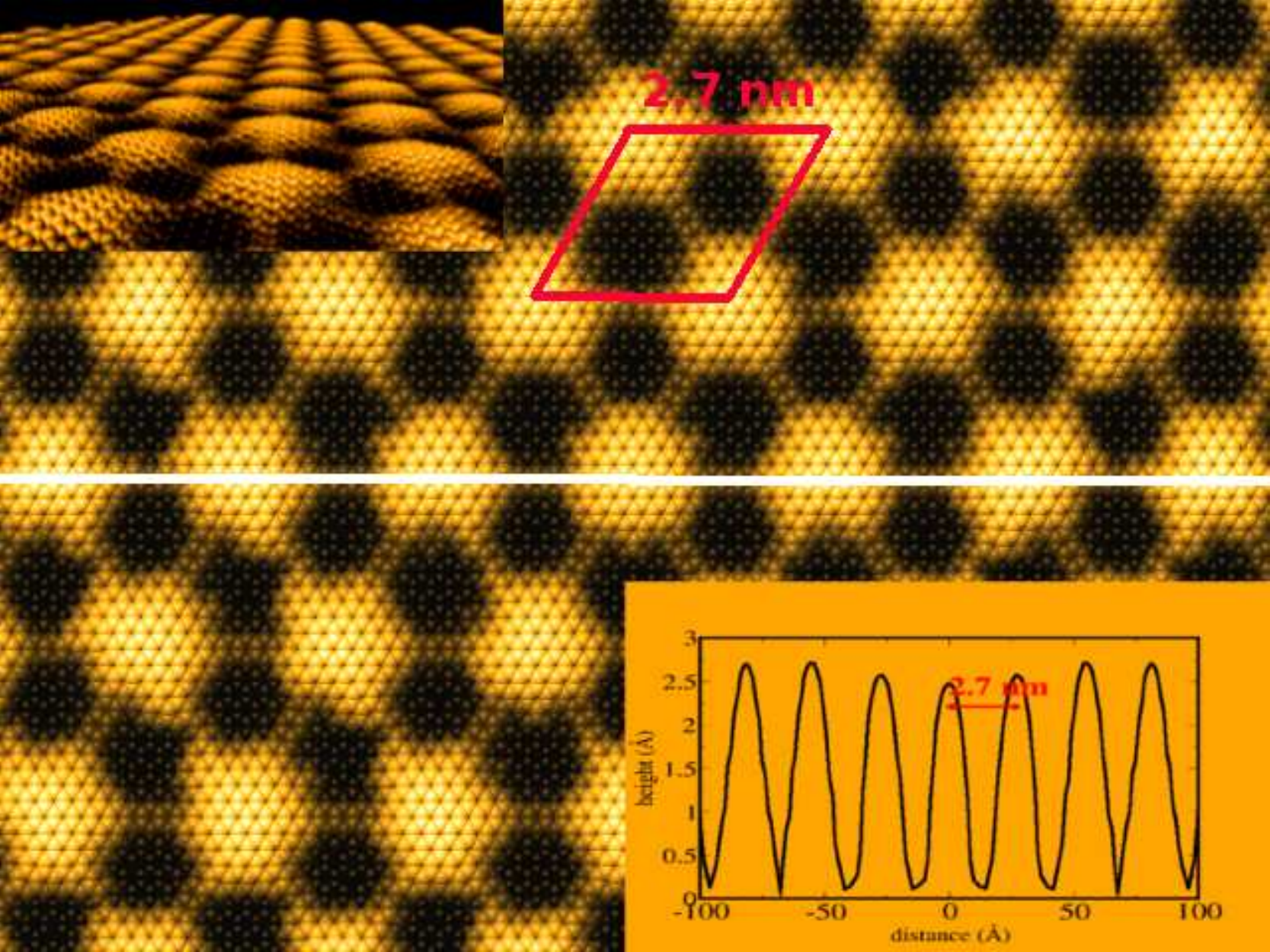}
\caption[]{
The gr/Ru(0001) system as obtained by classical molecular dynamics simulations
using the new Tersoff interfacial potential.
The height profiles are shown for gr (c) and for Ru(0001) (d).
The corrugations are $\xi=2.5$ $\hbox{\AA}$ (gr) and $\xi=0.2$ $\hbox{\AA}$ (Ru), respectively.
}
\end{center}
\label{F5}
\end{figure*}

\subsection{Results for the Abell-Tersoff interface}

 The moire superstructures obtained by the new Abell-Tersoff interface
are shown in Figs. 3.
(the minimal and the 
larger one) were cut out from a much larger
output structure (including few tens of thousands of C atoms) obtained from MD simulations 
using the newly developed C-Ru Tersoff potential.
In Figs. 3(a)-(b) the minimal rhomboid supercell is shown
($12 \times 12 / 11 \times 11$).
This minimal structure can be used to study adhesion and
corrugation by DFT \cite{DFT:Ru-Hutter,DFT:Ru-Stradi,DFT:Ru_corrug}.
In Fig. 3(c) the larger coincidence phase ($25 \times 25 /  23 \times 23$)  
(commensurate superstructure has been depicted which has 4 equivalent subunit \cite{Xray:Ru,LEED:Ru}.
The latter one has been experimentally seen by surface X-ray diffraction
(SXRD) approaches \cite{Xray:Ru} although the specific chiral
structure seen by SXRD \cite{Xray:Ru,SXRD:25X25} can not be identified in the presented Fig. 3(c).
The larger unit cell of the superstructure can be build up nearly
by 1000 C atoms which together with the Ru atoms presents
more than 2000 atoms which is beyond the reach of accurate DFT calculations
and geometry relaxation procedures and therefore only few studies
have been published on that yet \cite{LEED:Ru}.
In Fig. 3(d) the $12 \times 12$ supercell is shown embedded in a larger neighborhood
together with the Ru atoms below the gr-sheet. This figure together with Fig. 1(b) helps to
characterize in detail the most important binding sites and 
the notation of the adequate registry.

 In general, one can say that using a fitted Abell-Tersoff function
for the gr-metal couple, the overall landscape and topography
of the moir$\acute{e}$ superstructure changes significantly.
The bumps ruled pattern of the pure Morse or LJ potential dissapear and the correct "hump-and-bump" like features
can be seen. Hence, using bond order potentials
with angular dependence at the interface
the topography becomes essentially correct which is already comparable
with experimental STM results \cite{STM:Ru} (see Figs 3-4).
The most important properties of the supercells, the size and corrugation
are perfectly in accordance with experiment (supercell size)
or in reasonable agreement with many of the experimental and DFT methods.
In particular, the corrugation obtained by the Tersoff-only function
is in the range of $2-2.5$ $\hbox{\AA}$ at 300 K which is
somewhat larger than the average value obtained by DFT methods
($1.1-1.8$ $\hbox{\AA}$).
The overestimated corrugation is due to the cutoffed
C-Ru interactions at $3.2$ $\hbox{\AA}$.
The cutoff can not be avoided within the Abell-Tersoff formalism,
because above this value angular dependence becomes uncertain.

 It must be emphasized that this potential works only for the
corrugated gr-Ru(000) complex, and is inadequate for the
flat gr-Ru(000) system. 
This is because the angular term becomes sharply repulsive
for nonoptimal C-C-Ru bond angles ($\Theta \ne 80^{\circ} \pm 5$)
and the system transforms into the corrugated one.
At the flat gr-Ru(0001) interface $\Theta \approx 90^{\circ}$ and
adhesion becomes repulsive.
Hence the price what we pay is that the new potential is incorrectly
unstable for the flat supported graphene.
This price could still be tolerated if we consider the gain
what we obtain: the corrugated system is treated properly
and this new force field opens the way for large scale simulations
of supported graphene systems with moir\'e superstructures.
We see especially it useful e.g. for nanomechanical and
thermochemical simulations.
Moreover, we were unable to get perfectly or nearly flat gr on Ru(0001) either using
simple pairwise potentials (such as the Morse or LJ)
or Tersoff with other parameter set than used in this paper for
the corrugated structure.
Even if we use the AIREBO potential for Carbon \cite{airebo}
combined with LJ for the interface, buckling occurs with
a corrugation of $\sim 0.7$ $\hbox{\AA}$
although the out-of-plane movements are hindered in AIREBO
by a built-in torsional potential \cite{airebo} 
(off course the registry is wrong as it has been pointed out before
for pairwise interfacial potentials).
Hence, it seems that in the gr/Ru(0001) system there is a strong tendency
towards corrugation and flat gr does not persist on Ru(0001).
This is contrary to DFT calculations in which gr
remains nearly flat on Ru(0001) \cite{Peng}. In this calculations, however
the simple local density approximation has been used, which is known
from its notorious overbinding behaviour \cite{overbind} and which could flatten
gr unphisically.

 In Table III. the interfacial atomic distances are also depicted in which gr-topmost Ru(0001)
distances ($d_{min}$ and $d_{max}$) are shown. Within the humps (domes) 
$d_{max} \approx 3-5$ $\hbox{\AA}$ while at the bumps $d_{max} \approx 2.4 \pm 0.4$ $\hbox{\AA}$.
This is somewhat longer than found by DFT ($3.3$ and $2.2$).
Unfortunately, upon tuning the optimized potential we were not able to
decrease the distances.
The further decrease of the hump height requires the addition of
a long-range potential which could bind more effectively
the bulging regions.
The development of such a potential is, however, beyond the scope of the
present paper.


  The average bond distance in the bump regions $d \approx 2.1$ $\hbox{\AA}$
while in the humps at the peak $d \approx 4.0$ $\hbox{\AA}$.
This is consistent with those values given in DFT reports \cite{DFT:Ru-Hutter,DFT:Ru_corrug,DFT:Ru_Wang} in which $d_{if} \approx 2.1 \pm 0.1$ $\hbox{\AA}$
has been found in the flat bump regions and
$d_{if} \approx 3.8 \pm 0.1$ $\hbox{\AA}$ in the humps.
Experimental LEED and surface X-ray diffraction (SXRD)
studies report also $d_{if} \approx 2.1 \pm 0.1$ $\hbox{\AA}$ \cite{Batzill}
in the bulged in regions.

 The average coverage of the surface by physically sorbed surface features (humps)
can be estimated as $\phi_{cov} \approx 100.0 \times (1-E_{adh}/E_{chemical})
\approx 45 \%$ if we assume that
the adhesion energy is reduced due to the appearance of
non-bonding (or weakly) bonding regions (humps).
$E_{chemical}$ is the binding energy per Carbon for the chemically sorbed
regions (bumps).
The measured adhesion energy is the average of chemically and physically (vdw)
sorbed regions: $E_{adh}=(1-\phi_{cov})E_{chemical}+\phi_{cov} E_{vdw}$.
If we assume $E_{chemical} \approx 0.45$ eV/C \cite{DFT:Ru-Hutter,DFT:Ru-Stradi} and $E_{vdw} \approx 0.05$ eV/C
and $E_{adh} \approx 0.25$ eV/C,
then $\phi_{cov} \approx 0.55$.
Hence the larger part of the surface ($\sim 55 \%$) is covered by humps
(bulged regions) and which corresponds to hollow sites (hollow hump).


 The strain energy of the gr-sheets can also be calculated if we take
the difference of the Carbon atomic cohesive energies of the free relaxed
gr-sheet ($\sim -7.26$ eV/C) and the corrugated one (see Table III). 
The sheet strain is $E_{gr,str} \approx 0.15 \pm 0.04$ eV/C, which is
somewhat smaller than the gain in energy provided by adhesion.
DFT methods again provide 
$\sim 0.1$ eV/C \cite{DFT:Ru_Wang} which
includes contributions from bond stretching and other distortions (e.g. bending and torsional deformations) \cite{DFT:Ru_Wang}.
Interestingly, in few cases corrugated graphene proves to be more stable than
the flat graphene, hence the strain energy becomes negative (see Table III.,
hexagonal domes).
In particular, for the hexagonal domes we find that the strain energy
vs. the perfectly flat graphene is $0.02$ eV/C.
It is surprising that the hexagonal dome is very close in stability
to the flat gr, e.g. the energy difference is around thermal motion
at room temperature ($\sim 0.026$ eV).

 Restarting calculations just for the corrugated graphene layer using the
more sophisticated AIREBO Carbon potential \cite{airebo}
together with careful geometry relaxation and subsequent
0 K md simulation
we find $-7.80$ and $-7.78$ eV/C cohesive energies for the
perfectly flat and for the corrugated graphene, respectively.
This again suggests that the buckled gr-layer is very close in stability
to the plain sheet gr and the difference is within the magnitude of
thermal motion.

 The lattice misfit of the system has been effectively decreased from
the
initial $9.5 \%$ into $5.6 \%$ by leaving nearly unaltered the substrate's
lateral lattice constant (from $2.69$ to $2.686$ $\hbox{\AA}$) and
by stretching the sheet (from $2.46$ to $2.51$ $\hbox{\AA}$).
The serious increase of the gr-sheet lattice constant ($\sim 2 \%$)
could account mostly for the $\sim 0.11$ eV/C sheet strain energy which
is mostly accumulated at the hump regions.
Hence we conclude here that the gr-sheet is strongly
strained when placed on Ru(0001).
In ref. \cite{DFT:Ru_corrug} $\sim 3 \%$ C-C bond stretch has been mentioned
for regions with atop registries.
In ref. \cite{DFT:Ru_Wang} $\sim 1.8 \%$ is given.

\subsection{DFT geometry optimization of the corrugated and flat gr/Ru(0001) systems}

 {\em Ab initio} DFT geometry optimization (GO) has also been done
on a system with 748 atoms including the minimal rhomboid
supercell shown in Fig. 3(a).
The obtained corrugated structure can be seen in Fig. 6.
Structural details are given in Table III.
The initial structure for DFT GO has been cut out from CMD systems
with hexagonal domes.
If we initialize DFT GO from the CMD preoptimized
corrugated structure (with hexagonal domes) the buckled geometry persists.
Hence the CMD structure proved to be surprisingly well optimized for DFT
indicating that the developed force field works well.
It must also be noted that we get a small corrugation of $0.21$ $\hbox{\AA}$ if
the GO has been started from a flat gr on Ru(0001) using LDA approximation which is comparable with that of
given in ref. \cite{Peng} ($0.24$ $\hbox{\AA}$).
The GGA/PBE or vdW-DF2 functionals give the values of $0.23$ and $0.25$ $\hbox{\AA}$), respectively.
It has been pointed out by Peng {\em et al.} that the most likely periodic
boundary conditions applied at the cell border force the initially flat gr
sheet towards corrugation.
Therefore they used a quasiperiodic system, in which gr has been
treated as a noncontinous finite system and only the Ru substrate
has been treated as a periodic system.
In our simulations we also used non-periodic conditions to avoid
artificial constraints on corrugation provided by somewhat arbitrary
set periodicity (the proper choice of the lattice constant could be a difficult issue).
In non-periodic DFT/GO, however, the edge of the chosen system
could although influence the stability of the gr-flake
inducing edge-forces \cite{edgeforce},
however, a sufficiently large system could develop corrugation
realistically \cite{Peng}.
Starting from a flat gr system we get no ripples, except at the
edges, as it is expected \cite{edgeforce}.
The inner region remains nearly flat.
Hence edge-forces influence corrugation and induces edge-ripples
right at the borderline of the gr-flake \cite{edgeforce}.
Using
GO with periodic boundary conditions (PBCs) at the edges the corrugated
structure also persists and the buckled geometry becomes
preferential vs. periodic flat gr.
Therefore, excluding or at least minimizing edge-forces via 
PBCs graphene remains still seriously buckled.
It is another issue, as mentioned above, that PBCs could introduce
new "artifacts" on the structure, e.g., extra corrugation induced by
the periodic constraints \cite{Peng}.
\begin{figure}[hbtp]
\begin{center}
\includegraphics*[height=6cm,width=8cm,angle=0.]{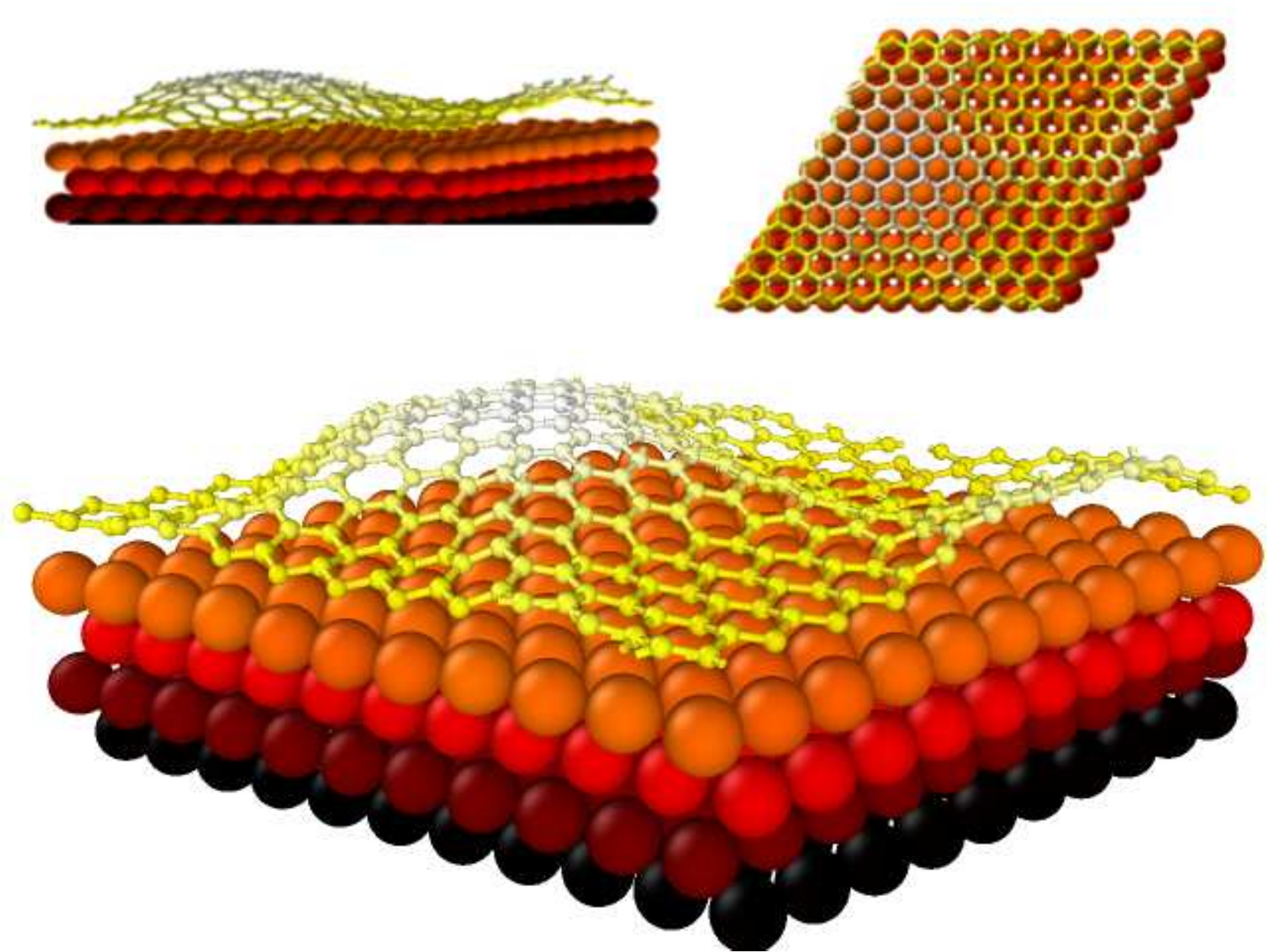}
\caption[]{
The buckled gr/Ru(0001) structure obtained by DFT/vdW-DF2 geometry optimization
starting from the hexagonal moire-corrugated CMD system shown
in Figs 4(c) and 5.
The initial systems have been cut out from the final relaxed CMD
structures with hexagonal domes.
The simulation cell corresponds to the minimal supercell shown
in Fig 3(a).
CMD simulations help to prepare initial moire-buckled structures
which would be otherwise hard to be obtained.
}
\end{center}
\label{F6}
\end{figure}
If the GO is initialized from a perfectly flat gr system, 
we get a weakly corrugated system (as already mentioned above) which is, however, largely stretched with the average C-C distance of $1.48$ $\hbox{\AA}$.
Our DFT/vdW-DF2 calculations find the nearly flat gr/Ru(0001) energetically
unfavorable. The strain energy (the energy/C vs. the fully relaxed perfectly flat gr) is $0.078$ eV/C) and the
adhesion energy is smaller ($-0.08$ eV/C) than in the corrugated system.
This result is partly in accordance with that of Peng {\em et al.} \cite{Peng}
who also found flat gr seriously stretched on Ru(0001).
Nevertheless, their results are somewhat different from ours
because they used the much simpler LDA approximation which is
well-known for its notorious overbinding tendency
and this behavior is related to the erroneous distance
dependence of approximate exchange functionals \cite{Harris}.
Due to this failure of LDA in describing weakly bound
systems they get a shorter C-Ru distance of $2.1$ $\hbox{\AA}$
than by the vdW-DF2 functional ($2.35$ $\hbox{\AA}$).
On the other hand, LDA performs surprisingly well for various
gr/support systems as far as certain phyisical properties are concerned (e.g. adhesion energy, bond distances) \cite{DFT_lda}.
Peng {\em et al.} also found the nearly flat gr strongly stretched ($1.43-1.49$ $\hbox{\AA}$
vs. our average values is $1.48$ $\hbox{\AA}$).
Our LDA GO gives similar results (results are not shown in Table III.)
to those of Peng {\em et al.} \cite{Peng}, however, CMD/LDA
as well as CMD/PBE also keep the corrugated structure obtained
by CMD.

 According to our calculations the energetical difference is
around $-0.075$ eV/atom with the vdW-DF2 functional ($-0.13$ eV/atom by LDA)
being the buckled one is more stable.
Since the difference is somewhat larger than the magnitude
of thermal motion $kT$ at room temperature ($\sim 0.026$ eV)
one can conclude that nearly flat gr is not competitive
on Ru(0001) vs. the corrugated one. 
Nevertheless, we do believe that the magnitude of this
difference is still small enough to consider both
corrugation regimes when experimental findings are to be
explained (the diversity of the measured corrugations).
In particular, this result could explain at least partly the experimental
controversy, that both low ($\xi < 0.5$ $\hbox{\AA}$) \cite{Parga,SXRD:25X25,STM:Ru} and highly ($0.5 < \xi < 1.8$ $\hbox{\AA}$) \cite{Xray:Ru,LEED:Ru,DFT:Ru_corrug,DFT:Ru_Wang,Giamfy} corrugated gr/Ru(0001)
systems have been reported.
Hence it could be that not only instrumental and other
experimental conditions result in the divergency of the measured
corrugation values but also gr can form different conformations
with a nearly equally stable structures.
It is, however, poorly understood yet which conformer is preferred by
gr under different experimental conditions.
Nevertheless, our DFT calculations predict the preferential formation of the
moire-corrugated system vs. the nearly flat conformer.

\section{Conclusions}

 Using classical molecular dynamics simulations we have shown
that it is possible to provide reasonably accurate results for weakly bound
systems such as gr/Ru(0001).
To achieve this task, however, it was important to develop
a new force field which adequately describe weak bonding between
the gr-sheet and Ru(0001).
It turned out that the widely used simple pair potentials (such as Lennard-Jones 
and Morse) lead to improper C-Ru bonding orientations and favor
incorrectly hollow-site registries instead of on-top ones.
The overall topography becomes then wrong: bumps (bulged-in regimes) can be found
at hollow-sites and the humps (bulges) at on-top registries.

 Properly oriented adhesion could only be reached with
a newly parameterized angular-dependent Abell-Tersoff potential for C-Ru interactions.
The obtained adhesion energy is very close to the DFT results.
The simulations automatically lead to the
$(12 \times 12)C/(11 \times 11)Ru$ nearly commensurate superstructures
and the obtained supercell size is in accordance with the available
experimental and DFT results. 
Larger coincidence supercells (e.g. $25on23$) can also be accounted for
by the new force field model.

 The application of the newly developed force-field could help to explain
the obtained STM micrographs. Moreover the Tersoff gr/Ru(0001) system
can also be used for various computer experiments such as
the simulation of the superior thermal conductivity of gr nanoribbons
or the nanomechanics of supported gr under external load.
In general, the new model opens the way towards large-scale supported gr-simulations
under various conditions
which was possible until recently only by simple pairwise potentials
which, however, provide inadequate moire structures as pointed
out in this article.

 We have also presented DFT calculations on a gr/Ru(0001) system corresponding
to the minimal conincidence supercell.
Surprisingly we find that when the DFT geometry optimization has been
initialized from the CMD/Tersoff corrugated structure with hexagonal domes
the moire-buckled geometry persists. 
Starting the DFT/GO from a flat system corrugation remains low ($\xi < 0.3$)
$\hbox{\AA}$. The energetic stability of the low and highly corrugated states
might be comparable, although the difference is slightly above the energy range of thermal motion $kT$ at room temperature (the moire-buckled system is more stable).
Therefore we conclude that despite the considerable strain energy in the strongly moire-corrugated
system the moire-buckled conformer of gr could be competitive or even more stable than the substrate supported flat gr.

%
%

\section{acknowledgement}
The calculations (simulations) have been
done mostly on the supercomputers
of the NIIF center (Hungary).
The kind help of P. Erhart (Darmstadt) in the usage of the PONTIFIX code is greatly acknowledged.
The availability of codes LAMMPS (S. Plimpton) and OVITO (A. Stukowski) are also greatly acknowledged.


\end{document}